\documentclass[prb,twocolumn,aps,superscriptaddress,floatfix]{revtex4-2}
\usepackage{amsmath,amssymb,amsbsy}
\usepackage{graphicx}
\usepackage{dcolumn}
\usepackage{bm}
\usepackage{xcolor}
\usepackage{epstopdf}
\usepackage{mathrsfs}
\usepackage{multirow}
\usepackage{ulem}
\usepackage{comment}
\usepackage[colorlinks,linkcolor=magenta,citecolor=blue,urlcolor=blue]{hyperref}
\usepackage{etoolbox}

\makeatletter
\patchcmd{\@makecaption}
{\scshape}
{}
{}
{}
\makeatother

\makeatletter
\newcommand{\Rmnum}[1]{\expandafter\@slowromancap\romannumeral #1@}
\makeatother

\begin{document}

\title{Limitations of SVD-Based Diagnostics for Non-Hermitian Many-Body Localization with Time-Reversal Symmetry}
\author{Huimin You}
\affiliation{Institute of Theoretical Physics and State Key Laboratory of Quantum Optics Technologies and Devices, Shanxi University, Taiyuan 030006, China}

\author{Jinghu Liu}
\affiliation{Department of Physics, Xinzhou Normal University, Xinzhou, Shanxi 034000, China}

\author{Yunbo Zhang}
\email{ybzhang@zstu.edu.cn}
\affiliation{Zhejiang Key Laboratory of Quantum State Control and Optical Field Manipulation, Department of Physics, Zhejiang Sci-Tech University, Hangzhou 310018, China}

\author{Zhihao Xu}
\email{xuzhihao@sxu.edu.cn}
\affiliation{Institute of Theoretical Physics and State Key Laboratory of Quantum Optics Technologies and Devices, Shanxi University, Taiyuan 030006, China}

\date{\today}

\begin{abstract}
Singular value decomposition (SVD) provides a convenient way to construct Hermitian-like diagnostics for non-Hermitian many-body systems, but its reliability for locating many-body localization (MBL) transitions remains unclear, particularly in systems preserving time-reversal symmetry (TRS). We benchmark SVD-based diagnostics against exact diagonalization (ED) in TRS-preserving non-Hermitian hard-core-boson chains with nonreciprocal hopping, considering quasiperiodic, random-disorder, and Stark potentials. We compare level statistics, half-chain entanglement entropy, inverse participation ratio, and spectral form factors. For the quasiperiodic and random-disorder models, ED-based entanglement and IPR yield mutually consistent finite-size transition estimates, whereas the corresponding SVD-based estimates are systematically shifted to larger disorder strengths and can lead to different phase assignments. The discrepancy is also reflected in the spectral form factors, where the ED-based dissipative spectral form factor and the SVD-based singular form factor can indicate different spectral regimes at the same parameters. In contrast, for the clean Stark model, ED and SVD give consistent transition estimates. We identify the origin of this model dependence as the fact that SVD probes the auxiliary Hermitian operator $\hat H^\dagger\hat H$, rather than the intrinsic right-eigenstate structure of $\hat H$; consequently, SVD can be quantitatively reliable only when the corresponding bulk-state structures remain aligned. Our results show that SVD-based diagnostics can capture qualitative RMT-to-Poisson trends, but are not generically reliable quantitative probes of MBL transitions in TRS-preserving non-Hermitian many-body systems.
\end{abstract}

\maketitle

\section{Introduction}

Many-body localization (MBL) is widely viewed as a paradigmatic mechanism by which an interacting system can avoid thermalization in the presence of strong disorder \cite{AbaninRMP2019,HuseP2016}. Unlike Anderson localization, which applies to noninteracting particles, MBL can remain stable even when interactions are present \cite{OganesyanPRB2007}. Consequently, the thermal-to-MBL transition is often discussed as a dynamical transition from an ergodic (thermalizing) regime to a localized, nonergodic regime. Reviews of this subject can be found in Refs.~\cite{NandkishoreARCMP2015,SierantRPP2025}, while representative studies include Refs.~\cite{BardarsonPRL2012,NandkishorePRB2014,GopalakrishnanPRB2016,KjallPRL2014}. Standard diagnostics in Hermitian systems include spectral and eigenstate properties: level statistics typically cross over from Wigner-Dyson to Poisson \cite{OganesyanPRB2007,PalPRB2010}, and eigenstate entanglement changes from volume-law scaling to area-law-like behavior \cite{BauerJSM2013}. A useful theoretical framework is provided by local integrals of motion (LIOMs), which explains the absence of transport and the persistence of memory at long times \cite{SerbynPapiPRL1102013,SerbynPapiPRL1112013,HuseNandkishorePRB2014}.

In recent years, there has been growing interest in extending these ideas to open quantum systems and non-Hermitian settings \cite{HamanakaPRB2025,RichterPRE2025}, where effective non-Hermitian Hamiltonians can arise, for example in quantum-trajectory dynamics conditioned on no quantum jumps \cite{DalibardPRL1992,CarmichaelPRL1993,PlenioRMP1998,DaleyAP2014,GongKawabataPRX2018,KawabataPRX2019}. A difficulty is that non-Hermitian Hamiltonians generally have complex eigenvalues, which complicates the use of standard spectral tools developed for Hermitian problems \cite{BenderPRL1998,HatanoNelsonPRL1996}; see Refs.~\cite{RotterJPAMT2009,AshidaAP2020}
for reviews and Refs.~\cite{WangSutharPRB2023,SutharPRB2025}
for recent many-body applications. To address this issue, several works have advocated the use of singular value decomposition (SVD), since singular values are always real and non-negative \cite{NandyPathakXianPRB2025,TekurPRB2024}. This makes it possible to define Hermitian-like diagnostics, such as singular-value spacing ratios \cite{KawabataPRXQ2023} and the singular form factor ($\sigma$FF), while keeping the numerical analysis straightforward. For instance, Ref. \cite{RoccatiPRB2024} applied singular-value statistics, $\sigma$FF, and properties of singular vectors to diagnose transitions between random-matrix theory (RMT)-like \cite{DysonJMP1401962,DysonJMP1571962,GuhrPR1998,BeenakkerRMP1997} and Poisson-like behavior in an interacting spin chain with random local dissipation. Notably, most SVD-based numerical studies have focused on non-Hermitian models that break time-reversal symmetry (TRS). In such TRS-breaking, class-A settings, previous studies have shown that SVD-based diagnostics can yield transition points that deviate from those obtained via ED, depending on the model and observables considered \cite{PrasadPRB2025}. Similar SVD-based analyses have also been applied to chaos-to-nonchaos crossovers in other TRS-breaking non-Hermitian models, including non-Hermitian Sachdev-Ye-Kitaev and random-matrix ensembles \cite{BaggioliPRD2025,NandyPathakTezukaPRB2025}. By contrast, the reliability of SVD diagnostics in TRS-preserving non-Hermitian many-body systems remains largely unexplored. Here TRS denotes an antiunitary symmetry $\mathcal{T}$ implemented as complex conjugation, satisfying $\mathcal{T}\hat{H}\mathcal{T}^{-1}=\hat{H}$ for the TRS-preserving models considered below. We therefore ask whether SVD-based diagnostics can faithfully locate the MBL transition in TRS-preserving non-Hermitian systems.

TRS can play a crucial role in non-Hermitian many-body systems, but it has been less emphasized in the existing SVD-based literature. In disordered non-Hermitian systems preserving TRS, the spectrum often exhibits a real-to-complex transition that is frequently discussed in connection with the MBL transition \cite{HamazakiPRL2019,ZhaiPRB2020}. By contrast, when TRS is broken, such a spectral transition is generally absent, even if MBL persists \cite{HamazakiPRL2019}. However, TRS alone does not ensure that spectral and MBL transitions coincide: in the TRS-preserving interacting Stark model, the Stark-MBL transition and the real-to-complex spectral transition occur at parametrically different tilt strengths \cite{LiuXuPRB2023,LiYuPRA2023}. These observations motivate us to examine, in TRS-preserving non-Hermitian systems, when SVD-based diagnostics are consistent with ED-based indicators of the MBL transition, and to delineate the regimes in which they may deviate.

To this end, we perform a side-by-side comparison between ED- and SVD-based probes, including spacing ratios \cite{SakhrPRE2006,AtasJPAMT2013,CorpsPRE2020,TekurPRB2018}, half-chain entanglement entropy \cite{KumarPRB2023,TianPRB2025}, inverse participation ratio (IPR) \cite{EversPRL2000,SutharWangPRB2022}, and spectral form factors (SFFs) \cite{BertiniPRL2018,WeiPRE2024,BrezinPRE1997}. We study three representative TRS-preserving non-Hermitian hard-core-boson chains with quasiperiodic, random-disorder, and Stark potentials. This comparison allows us to identify parameter regimes where the two approaches agree or disagree in classifying ergodic-like versus MBL-like behavior.

\section{Model and Diagnostic Methods}

In this work, we combine ED with SVD to systematically assess the range and limitations of SVD-based diagnostics in TRS-preserving non-Hermitian many-body systems. We study three representative hard-core-boson chains subject to (i) a quasiperiodic potential, (ii) random disorder, and (iii) a Stark potential. In all three settings, the Hamiltonian can be written as $\hat{H}=\hat{H}_I+\hat{W}$, where
\begin{equation}
  \hat{H}_{I}
  =\sum_{j}\left[-J\left(e^{-g} \hat{b}_{j+1}^{\dagger}\hat{b}_{j} + e^{g} \hat{b}_{j}^{\dagger}\hat{b}_{j+1}\right) +U\hat{n}_{j}\hat{n}_{j+1}\right],
  \label{eq:main_hamiltonian}
\end{equation}
and $\hat{W}=\sum_j W_j\hat{n}_j$ encodes the model-dependent on-site potential. For all models considered in the main text, $J$, $U$, $g$, and $W_j$ are real. Therefore, in the occupation-number basis the Hamiltonian matrix is real and satisfies $\mathcal T \hat H \mathcal T^{-1}=\hat H$ with $\mathcal T=\mathcal K$, where $\mathcal K$ denotes complex conjugation. Unless otherwise specified, the hopping term is implemented with periodic boundary conditions. For the Stark model, the on-site potential is defined with respect to a fixed spatial origin on the finite lattice, following standard finite-size Stark-MBL calculations. Here, $\hat{b}_{j}$ ($\hat{b}_{j}^{\dagger}$) annihilates (creates) a hard-core boson on site $j$, and $\hat{n}_{j}=\hat{b}_{j}^{\dagger}\hat{b}_{j}$ is the local number operator. The parameter $J$ sets the hopping amplitude, and $U$ is the nearest-neighbor interaction strength. Non-Hermiticity is introduced via nonreciprocal hopping controlled by $g$. For $g\neq 0$, left and right hopping amplitudes differ by factors $e^{\mp g}$. We directly compare ED- and SVD-based indicators at the level of both eigenstate properties and spectral correlations. This comparison allows us to identify regimes where SVD diagnostics agree or disagree with ED-based indicators of the MBL transition. Within ED, the $n$th eigenvalue and the corresponding right eigenstate are denoted by $E_n$ and $|\psi_n\rangle$, respectively. In the SVD framework, $\hat{H}=U\Sigma V^{\dagger}=\sum_{n}\sigma_{n}|u_{n}\rangle\langle v_{n}|$, where $U=\left(|u_{1}\rangle,\ldots,|u_{D}\rangle\right)$ and $V=\left(|v_{1}\rangle,\ldots,|v_{D}\rangle\right)$ are the left and right singular vectors, respectively, and $\Sigma=\mathrm{diag}\left(\sigma_{1},\ldots,\sigma_{D}\right)$ contains the singular values. We denote the $n$th singular value and the corresponding right singular vector by $\sigma_n$ and $|v_n\rangle$, respectively. We emphasize that the right singular vectors are eigenvectors of the auxiliary Hermitian operator $\hat H^\dagger \hat H$, rather than eigenvectors of $\hat H$ itself. For a generic non-normal Hamiltonian satisfying $[\hat H,\hat H^\dagger]\neq 0$, these two sets of vectors need not encode the same spatial or entanglement structure. This distinction is central to the ED--SVD comparison below. Unless stated otherwise, we work at half filling $N/L=1/2$, where $N$ is the particle number and $L$ is the system size; the corresponding Hilbert-space dimension is $D=\binom{L}{N}$. For cases (i) and (ii), spacing ratios and eigenstate-based quantities are averaged over disorder realizations and, for each realization, over a fixed bulk spectral window; throughout, the bulk window is taken to be the central one-fifth of the spectrum. For the complex ED spectrum, this central window is selected by ordering eigenvalues according to their distance from the spectral center in the complex-energy plane, while for the singular-value spectrum it is selected from the central part of the ordered singular values. For the Stark case (iii), where no disorder ensemble is present, we restrict these local spectral and eigenstate averages to the same bulk window only.

We characterize spectral correlations using the level-spacing ratio, a standard diagnostic in quantum chaos and RMT that quantifies level repulsion and thus distinguishes RMT-like from Poisson-like spectral correlations. For the complex spectrum obtained from ED, we employ the complex spacing ratio \cite{PrasadPRB2025,SutharWangPRB2022,SaPRX2020,YusipovAIJNS2022}
\begin{equation}
  z_n=\frac{E_n-E_{\mathrm{NN}}}{E_n-E_{\mathrm{NNN}}}=\left|z_n\right|e^{i\varphi_n},\label{eq:complex_spacing_ratio}
\end{equation}
where $E_{\mathrm{NN}}$ and $E_{\mathrm{NNN}}$ are the nearest and next-nearest neighbors of $E_n$ in the complex-energy plane, respectively. Here ``nearest'' is determined by the Euclidean distance $|E_m-E_n|$ in the complex plane, and $|z_n|$ and $\varphi_n$ are the magnitude and argument of $z_n$. We define the disorder- and bulk-spectrum-averaged radial statistic as $\langle r^{\text{E}}\rangle=\langle |z|\rangle$, where $\langle\cdots\rangle$ indicates an average over eigenvalues in the chosen bulk window and over disorder realizations. In non-Hermitian systems, $\langle r^{\text{E}}\rangle$ typically approaches the Ginibre (RMT-like) value $\langle r^{\text{E}}\rangle \approx 0.74$ in regimes with strong spectral correlations, whereas it tends to the Poisson value $\langle r^{\text{E}}\rangle \approx 0.5$ when spectral correlations are weak \cite{SutharWangPRB2022,SaPRX2020,PeronPRE2020}. Within the SVD framework, the singular values form a real, non-negative spectrum. We order them in ascending order, and use the conventional spacing ratio \cite{OganesyanPRB2007,WangSunPRB2021}
\begin{equation}
  r_{n}=\frac{\min \left( \delta _{n+1},\delta _{n}\right) }{\max \left(
    \delta _{n+1},\delta _{n}\right) },  \label{eq3}
\end{equation}
with the nearest-neighbor spacing $\delta _{n}=\sigma_{n+1}-\sigma_{n}$. The corresponding disorder- and bulk-spectrum average is denoted by $\langle r^{\text{S}}\rangle =\langle r \rangle$ \cite{RoccatiPRB2024,PrasadPRB2025}. When the singular-value spectrum exhibits RMT-like correlations, $\langle r^{\mathrm{S}}\rangle$ approaches the Gaussian orthogonal ensemble (GOE) value $\langle r^{\mathrm{S}}\rangle \approx 0.53$, while it crosses over to the Poisson value $\langle r^{\mathrm{S}}\rangle \approx 0.39$ in the weakly correlated regime \cite{OganesyanPRB2007,AtasPRL2013}. Comparing $\langle r^{\mathrm{E}}\rangle$ and $\langle r^{\mathrm{S}}\rangle$ allows us to diagnose whether spectral correlations are RMT-like or Poisson-like, while the nature of the underlying eigenstates is assessed in conjunction with entanglement entropy and IPR. In the main text we focus on the radial component $|z_n|$, because it gives a rotationally averaged measure of local level repulsion and is sufficient for identifying the RMT-to-Poisson crossover relevant to our ED-based spectral diagnosis. The angular component $\varphi_n$ can provide additional information about spectral anisotropy, clustering near the real axis, and real-to-complex spectral evolution. However, it has no direct analogue in the SVD-based singular-value statistics, since singular values are nonnegative real numbers. Therefore, angular statistics are complementary to, rather than central for, the ED--SVD comparison considered here. We analyze the angular component in Appendix~\ref{app:angular_statistics}, where we show that it does not qualitatively change the conclusions drawn from the radial statistic.

Entanglement entropy provides a key quantity for characterizing the MBL transition. We consider the half-chain entanglement entropy of the $n$th state \textcolor{red}{\cite{WangSutharPRB2023,SutharWangPRB2022,ZhangZhangPRB2020}}
\begin{equation}
  S_n=-\mathrm{Tr}\left( \rho^n_{A}\ln \rho^n_{A}\right),  \label{eq4}
\end{equation}
where the system is bipartitioned into two equal subsystems $A$ and $B$ with $L_A=L_B=L/2$. The reduced density matrix $\rho_A^{n}=\mathrm{Tr}_B(\rho^{n})$ is obtained by tracing out subsystem $B$. In the ED approach, we construct the density matrix from the $n$th right eigenstate as $\rho^{n}=|\psi_n\rangle\langle\psi_n|$. In the SVD approach, we analogously define $\rho^{n}=|v_{n}\rangle\langle v_{n}|$ using the $n$th right singular vector $|v_n\rangle$ \cite{RoccatiPRB2024}. We denote the disorder- and bulk-spectrum-averaged half-chain entanglement entropies extracted from ED and SVD by $\langle S^{\mathrm{E}}\rangle$ and $\langle S^{\mathrm{S}}\rangle$, respectively. The scaling of entanglement entropy with system size distinguishes extended and localized eigenstates \cite{SuntajsPRB2020}. In extended/ergodic regimes, the entanglement entropy exhibits volume-law scaling, consistent with highly entangled states. In localized/MBL-like regimes, it approaches an area-law-like scaling, reflecting reduced entanglement \cite{SerbynPapiPRL1112013,WangSutharPRB2023}.

Complementary to spectral and entanglement diagnostics, we further characterize the spatial distribution properties of the states via the IPR, which quantifies how localized a normalized state is in a chosen basis. For a Hilbert space dimension $D$, the IPR of the $n$th state is defined as \cite{SerbynPapiPRL1112013,SutharWangPRB2022,ModakPRL2015}
\begin{equation}
  \mathrm{IPR}_{n}=\sum_{k=1}^{D}\left\vert c_{k}\right\vert ^{4},  \label{eq5}
\end{equation}
where $c_k$ are the expansion coefficients in the computational basis $\{|e_k\rangle\}$. In the ED approach, $c_k^{\mathrm{E}}=\langle e_k|\psi_n\rangle$, while in the SVD approach we analogously define $c_k^{\mathrm{S}}=\langle e_k|v_n\rangle$ using the right singular vector $|v_n\rangle$ \cite{RoccatiPRB2024,PrasadPRB2025}. The IPR distinguishes localized and extended structures: in a strongly localized phase, $\mathrm{IPR}_n=O(1)$, whereas for a fully extended state one expects $\mathrm{IPR}_n=O(1/D)$ \cite{EversPRL2000,ModakPRL2015,RoccatiPRB2024}. Accordingly, we denote the disorder- and bulk-spectrum-averaged IPR obtained from ED and SVD by $\langle \mathrm{IPR}^{\mathrm{E}}\rangle$ and $\langle \mathrm{IPR}^{\mathrm{S}}\rangle$, respectively.

We also study SFFs, which probe long-range spectral correlations in the time domain and provide a sensitive diagnostic of RMT-like versus Poisson-like behavior. For the complex energy spectrum $\{E_n\}$ obtained from ED, we use the dissipative SFF (DSFF) \cite{LiProsenPRL2021,GhoshPRB2022}
\begin{equation}
  \mathrm{DSFF}(\tau,\tau^\ast)  =  \left|\frac{1}{D}\sum_{n=1}^{D}  \exp\left[\frac{i}{2}\left(E_n\tau^*+E_n^*\tau\right)\right]\right|^2,
  \label{eq6}
\end{equation}
where $\tau= t e^{i\theta}$ defines a complex ``time'' variable. $\tau$ is a Fourier-conjugate parameter to the real and imaginary parts of the complex spectrum rather than the physical real time of unitary dynamics. We take $\theta\in[0,\pi/2]$ and fix $\theta=\pi/4$ throughout, which treats the real and imaginary parts on equal footing and is appropriate for generic complex spectra. The robustness of the DSFF results against other choices of $\theta$ is checked for both the quasiperiodic and random-disorder models in Appendix~\ref{app:dsff_angle}. For an uncorrelated complex spectrum, the Poisson reference under the normalization of Eq.~(\ref{eq6}) is
\begin{equation}
  \mathrm{DSFF}_{\mathrm{Poisson}}(\tau,\tau^\ast)
  =
  \frac{1}{D}
  +
  \frac{D-1}{D}e^{-|\tau|^2},
  \label{eq:dsff_poisson}
\end{equation}
which approaches the plateau value $1/D$ at long times \cite{LiProsenPRL2021}.

Within the SVD approach, we consider the $\sigma$FF defined in terms of the unfolded singular values $\{\tilde{\sigma}_n\}$, which are obtained by mapping the singular values of the non-Hermitian Hamiltonian through a standard unfolding procedure based on a smooth fit to the cumulative distribution function \cite{HaakeS2010}, thereby ensuring that global density variations do not contaminate the correlation measures \cite{RoccatiPRB2024,NandyPathakTezukaPRB2025,CotlerJHEP2017,LiuPRD2018,LozejE2023,BianchiJHEP2024,SuntajsPRE2020,PrakashPRR2021}
\begin{equation}
  \sigma\mathrm{FF}(t)
  =
  \left|\frac{1}{D}\sum_{n=1}^{D}e^{-i\tilde{\sigma}_n t}\right|^2.
  \label{eq9}
\end{equation}
This quantity plays a role analogous to the conventional SFF in Hermitian systems, but is constructed from the real, non-negative singular spectrum. In regimes with RMT-like spectral correlations, DSFF and $\sigma\mathrm{FF}(t)$ exhibit the characteristic dip--ramp--plateau structure and approach the corresponding RMT predictions. A commonly used GOE benchmark, normalized to approach $1/D$ at late times, is \cite{LiuPRD2018,LozejE2023,BianchiJHEP2024,SuntajsPRE2020}
\begin{equation}
  \sigma\mathrm{FF}_{\mathrm{GOE}}(s)=
  \begin{cases}
    \frac{2s}{D}-\frac{s}{D}\ln\!\left(1+2s\right),              & s\le 1, \\
    \frac{2}{D}-\frac{s}{D}\ln\!\left(\dfrac{2s+1}{2s-1}\right), & s\ge 1,
  \end{cases}
  \label{eq:goe_sigmaff}
\end{equation}
where $s=t/t_{\mathrm{H}}$ is the dimensionless time rescaled by the Heisenberg time $t_{\mathrm{H}}$, set by the mean level spacing of the unfolded singular-value spectrum. In Poisson-like regimes, the ramp is strongly suppressed. Therefore, together with level spacing ratios and eigenstate-based measures, DSFF and $\sigma$FF provide additional diagnostics for distinguishing RMT-like from Poisson-like behavior.

In this paper, we set $J=1$ as the energy unit and fix $U=2$ and, unless stated otherwise, $g=0.5$. For both the quasiperiodic and random models, we average over $8000$, $3000$, and $200$ independent realizations for $L=10$, $12$, and $14$, respectively. The finite-size scaling protocol used to extract the quoted transition estimates is described in Appendix~\ref{app:scaling}.

\section{Quasiperiodic Model}

\begin{figure}[t]
  \includegraphics[width=\linewidth]{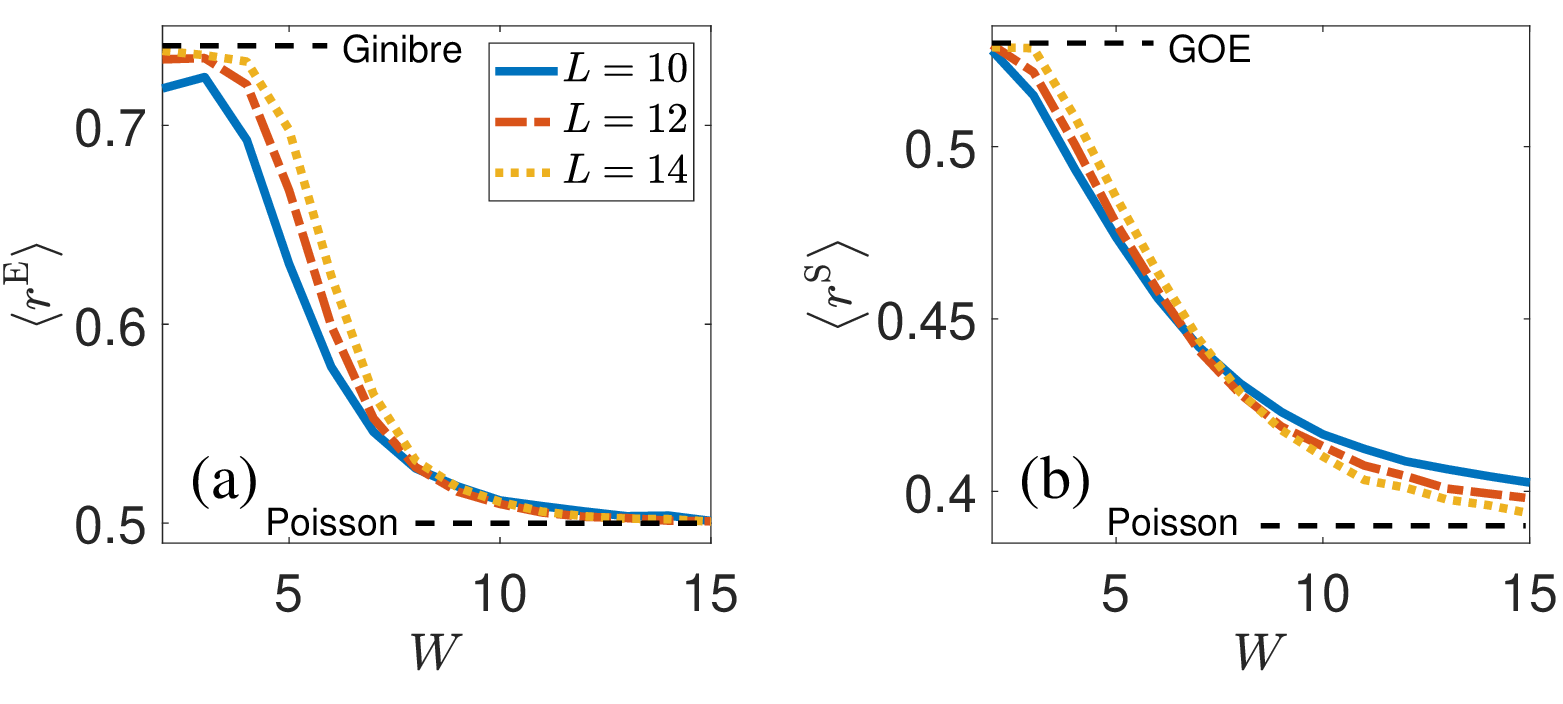}
  \caption{Disorder-averaged spacing ratios (a) $\langle r^{\mathrm{E}}\rangle$ and (b) $\langle r^{\mathrm{S}}\rangle$ versus quasiperiodic potential strength $W$ for different system sizes. Dashed lines indicate the corresponding RMT and Poisson reference values.}
  \label{FIG1}
\end{figure}

\begin{figure}[t]
  \includegraphics[width=\linewidth]{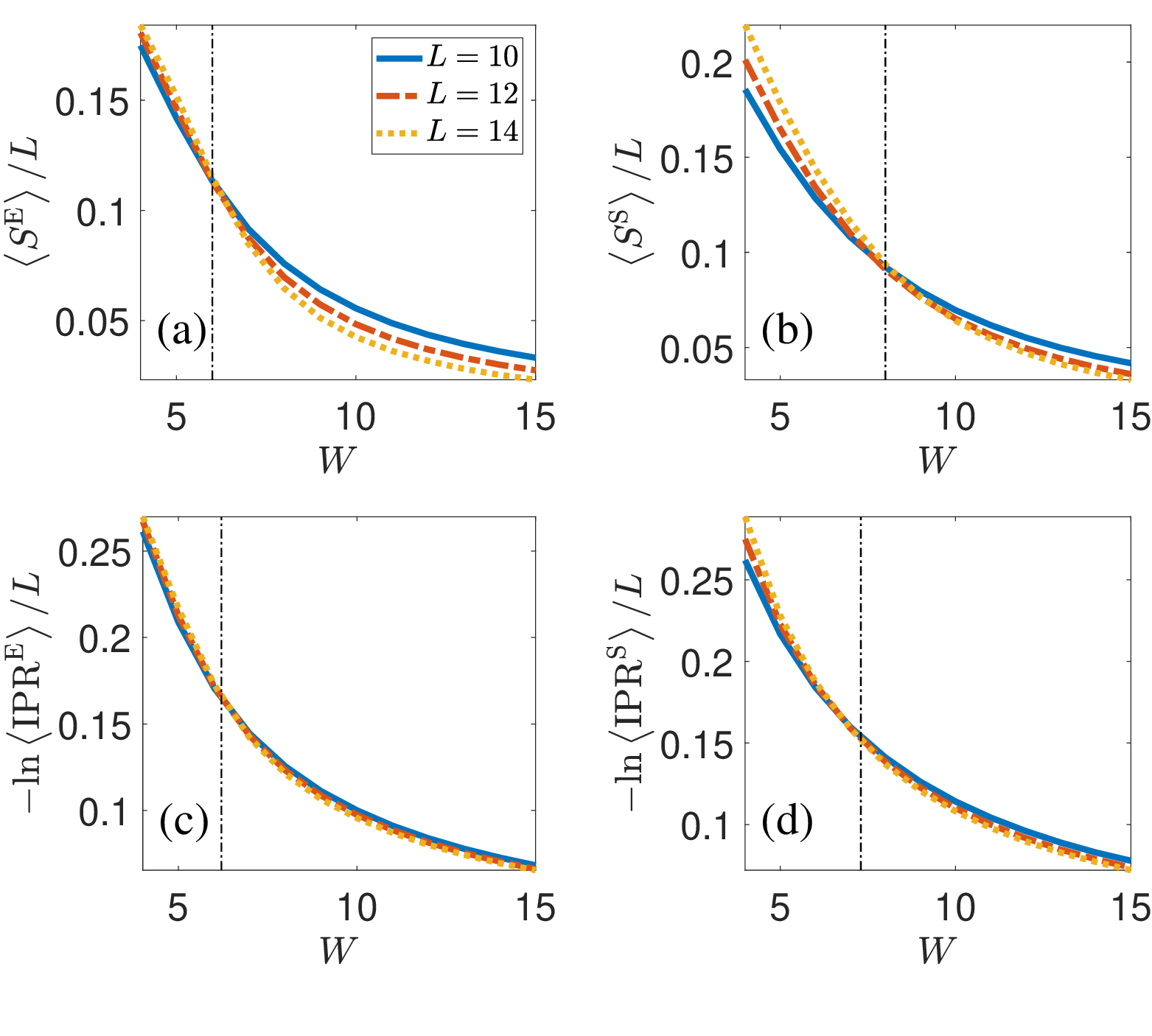}
  \caption{Half-chain entanglement entropy and IPR versus quasiperiodic strength $W$ for different system sizes. Entanglement: (a) $\langle S^{\mathrm{E}}\rangle$ and (b) $\langle S^{\mathrm{S}}\rangle$. IPR: (c) $\langle \mathrm{IPR}^{\mathrm{E}}\rangle$ and (d) $\langle \mathrm{IPR}^{\mathrm{S}}\rangle$. Dashed lines mark the finite-size transition estimates obtained from the scaling analysis.}
  \label{FIG2}
\end{figure}

\begin{figure}[t]
  \includegraphics[width=\linewidth]{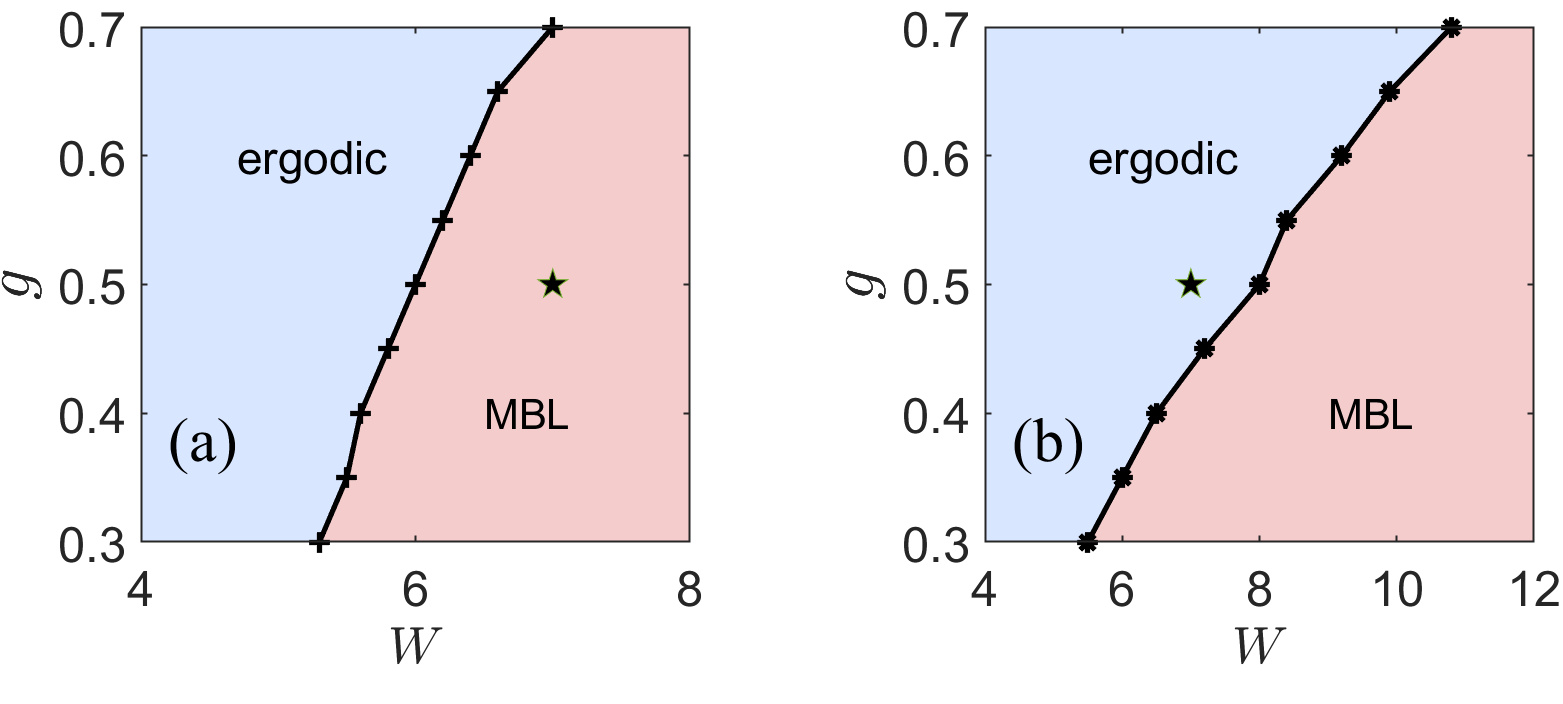}
  \caption{Phase diagram of the quasiperiodic model in the $(g,W)$ plane. (a) ED boundary extracted from $\langle S^{\mathrm{E}}\rangle$. (b) SVD boundary extracted from $\langle S^{\mathrm{S}}\rangle$. The marked point indicates the representative parameters $(g,W)=(0.5,7)$.}
  \label{FIG3}
\end{figure}

We consider a TRS-preserving non-Hermitian hard-core-boson chain subject to a quasiperiodic on-site potential, $W_{j}=W\cos \left( 2\pi \alpha j+\phi \right)$, where $W$ is the potential strength and $\phi\in[0,2\pi)$ is a random phase. We choose an irrational modulation wave number $\alpha=(\sqrt{5}-1)/2$.

To diagnose an ergodic-to-MBL transition, we compare spectral-statistics diagnostics obtained from ED and SVD. We first evaluate the level-spacing ratio defined in Eq.~(\ref{eq:complex_spacing_ratio}) using the complex many-body eigenvalues. As shown in Fig.~\ref{FIG1}(a), $\langle r^{\mathrm{E}}\rangle$ is close to the Ginibre value at weak quasiperiodic potential and gradually drifts toward the Poisson limit as $W$ increases, consistent with a finite-size crossover from an ergodic-like regime to an MBL-like regime. We then repeat the analysis in the SVD framework by computing the spacing ratio defined in Eq.~(\ref{eq3}) for the singular values. Figure~\ref{FIG1}(b) shows a similar drift of $\langle r^{\mathrm{S}}\rangle$ from an RMT-like value toward the Poisson limit.

To locate the transition more accurately, we further analyze the half-chain entanglement entropy. In the ergodic regime, the entanglement entropy exhibits volume-law scaling, whereas in the MBL regime it crosses over to an area-law-like behavior \cite{SerbynPapiPRL1112013}. Concretely, we compute the random-phase- and bulk-spectrum-averaged half-chain entanglement entropies $\langle S^{\mathrm{E}}\rangle$ and $\langle S^{\mathrm{S}}\rangle$ for three system sizes, with the results shown in Fig.~\ref{FIG2}(a) and Fig.~\ref{FIG2}(b), respectively. The finite-size scaling collapse in Fig.~\ref{FIG2}(a) yields an ED-based transition estimate $W_{c}^{\mathrm{E}}\approx 6$. By contrast, the corresponding collapse in Fig.~\ref{FIG2}(b) places the SVD-based transition at a substantially larger value, $W_{c}^{\mathrm{S}}\approx 8$. This systematic shift indicates that, in this TRS-preserving quasiperiodic non-Hermitian setting, the SVD-based entanglement diagnostic does not reliably track the ED-inferred transition point.

As an independent check, we compute the IPR for both approaches. Specifically, we evaluate $\langle \mathrm{IPR}^{\mathrm{E}}\rangle$ and $\langle \mathrm{IPR}^{\mathrm{S}}\rangle$ for the same three system sizes, which are shown in Fig.~\ref{FIG2}(c) and Fig.~\ref{FIG2}(d). Consistent with the entanglement analysis, both quantities display a size-dependent crossover toward stronger localization as $W$ increases. The finite-size scaling collapse yields $W_{c}^{\mathrm{E}}\approx 6.2$ from $\langle \mathrm{IPR}^{\mathrm{E}}\rangle$ and $W_{c}^{\mathrm{S}}\approx 7.3$ from $\langle \mathrm{IPR}^{\mathrm{S}}\rangle$. The persistence of an ED--SVD mismatch across two independent observables suggests that the discrepancy is not an artifact of a particular diagnostic, but rather reflects that singular-vector-based measures are not reliable for quantitatively locating the transition in this TRS-preserving quasiperiodic non-Hermitian model.

To provide an intuitive overview, we summarize these results in the phase diagrams shown in Fig.~\ref{FIG3}. Figure~\ref{FIG3}(a) displays the ED-based finite-size phase diagram extracted from the half-chain entanglement entropy, where the solid line marks the estimated crossover boundary separating an ergodic-like regime from an MBL-like regime in the $(g,W)$ plane. Figure~\ref{FIG3}(b) presents the corresponding SVD-based finite-size phase diagram obtained from the same entanglement criterion. Although the two phase diagrams share a similar overall structure, the SVD-based boundary is systematically shifted to larger $W$. Importantly, this shift can lead to a qualitative misclassification of the phase at the same point in parameter space. As a representative example, the point $(g,W)=(0.5,7)$ in Fig.~\ref{FIG3} lies on the localized side of the ED-based boundary, while it is still identified as ergodic by the SVD-based boundary, directly highlighting that the SVD approach is not reliable for determining the phase boundary in this TRS-preserving quasiperiodic non-Hermitian setting.

\begin{figure}[t]
  \includegraphics[width=\linewidth]{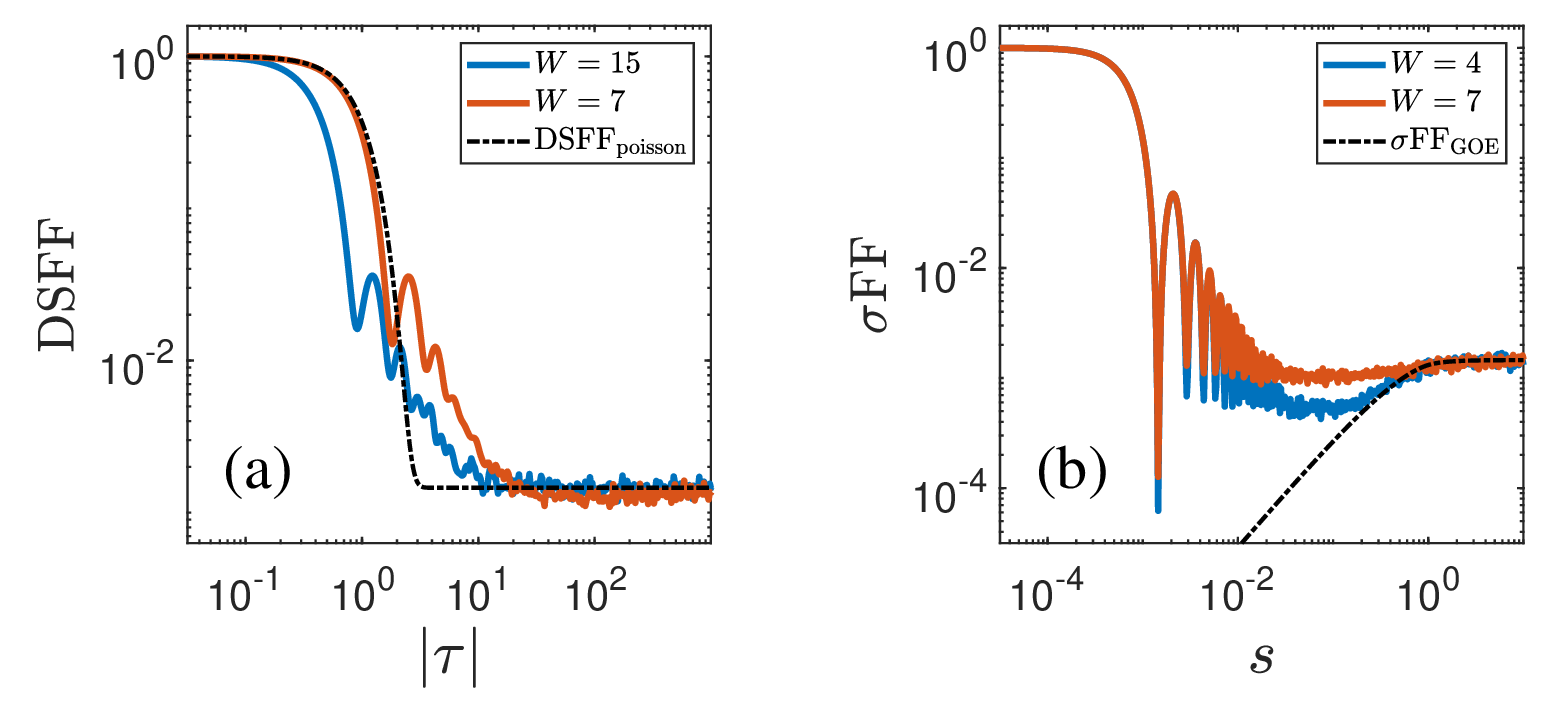}
  \caption{Dynamics of the DSFF and $\sigma$FF for the quasiperiodic model at $L=14$, averaged over 200 realizations. (a) DSFF for $W=7$ and $15$; the dashed line marks the Poisson plateau $1/D$. (b) $\sigma$FF for $W=4$ and $7$; dashed lines show the GOE predictions.}
  \label{FIG4}
\end{figure}

Finally, we employ DSFF and $\sigma$FF to probe the spectral correlations in the time domain across the transition region, as shown in Figs.~\ref{FIG4}(a) and \ref{FIG4}(b), respectively. Both quantities are computed for $L=14$ and averaged over $200$ realizations of the random phase $\phi$. In Fig.~\ref{FIG4}(a), the DSFF at $W=7$ and $W=15$ shows no discernible ramp, indicating Poisson-like behavior. In contrast, Fig.~\ref{FIG4}(b) shows that the $\sigma$FF at $W=4$ exhibits the characteristic dip--ramp--plateau structure \cite{RoccatiPRB2024,NandyPathakTezukaPRB2025}, indicating level repulsion consistent with RMT and hence an ergodic phase. At the intermediate value $W=7$, a ramp is still visible in $\sigma$FF but is suppressed, while the DSFF remains Poisson-like. This difference between DSFF and $\sigma$FF at the same parameter point provides dynamical support that in TRS-preserving quasiperiodic non-Hermitian systems, SVD-based diagnostics can yield inconsistent transition classifications and are therefore not reliable for pinpointing the ergodic-to-MBL transition.

\section{Random-disorder Model}

\begin{figure}[t]
  \includegraphics[width=\linewidth]{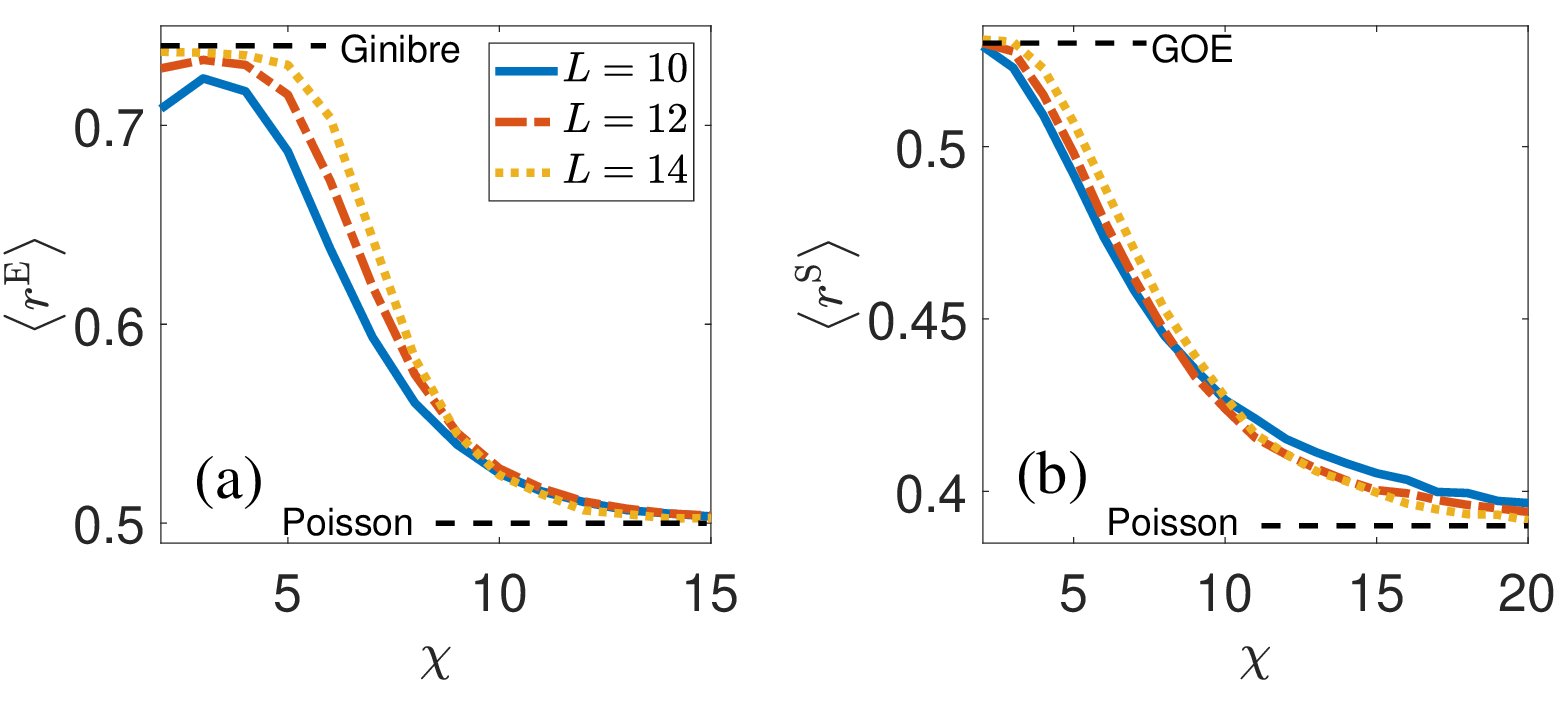}
  \caption{Disorder-averaged spacing ratios (a) $\langle r^{\mathrm{E}}\rangle$ and (b) $\langle r^{\mathrm{S}}\rangle$ versus random disorder strength $\chi$ for different system sizes $L$. Dashed lines denote the corresponding RMT and Poisson reference values.}
  \label{FIG5}
\end{figure}

\begin{figure}[t]
  \includegraphics[width=\linewidth]{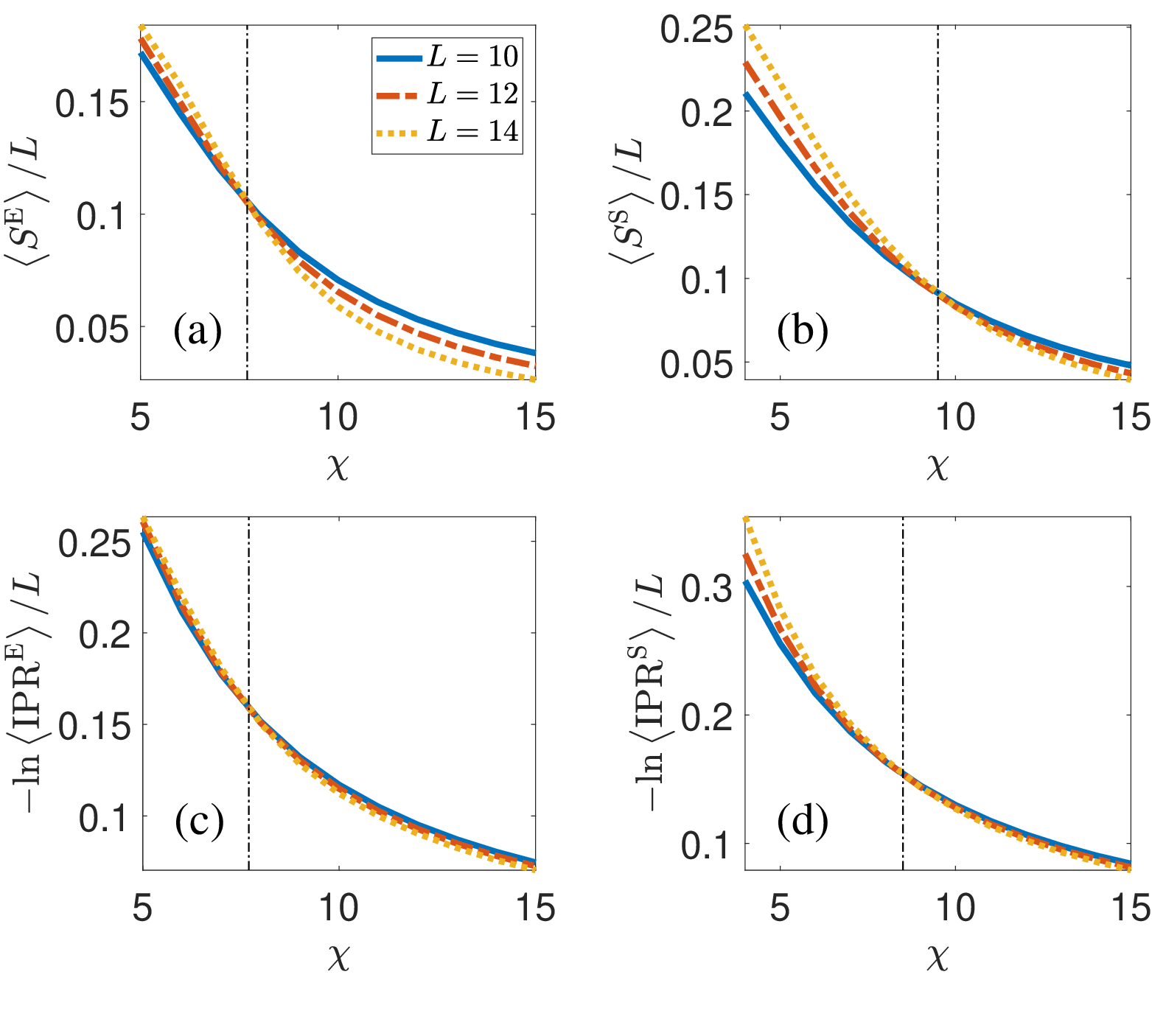}
  \caption{Half-chain entanglement entropy and IPR versus random disorder strength $\chi$ for different system sizes $L$. Entanglement: (a) $\langle S^{\mathrm{E}}\rangle$ and (b) $\langle S^{\mathrm{S}}\rangle$. IPR: (c) $\langle \mathrm{IPR}^{\mathrm{E}}\rangle$ and (d) $\langle \mathrm{IPR}^{\mathrm{S}}\rangle$. Dashed lines mark the finite-size transition estimates obtained from the scaling analysis.}
  \label{FIG6}
\end{figure}

\begin{figure}[t]
  \includegraphics[width=\linewidth]{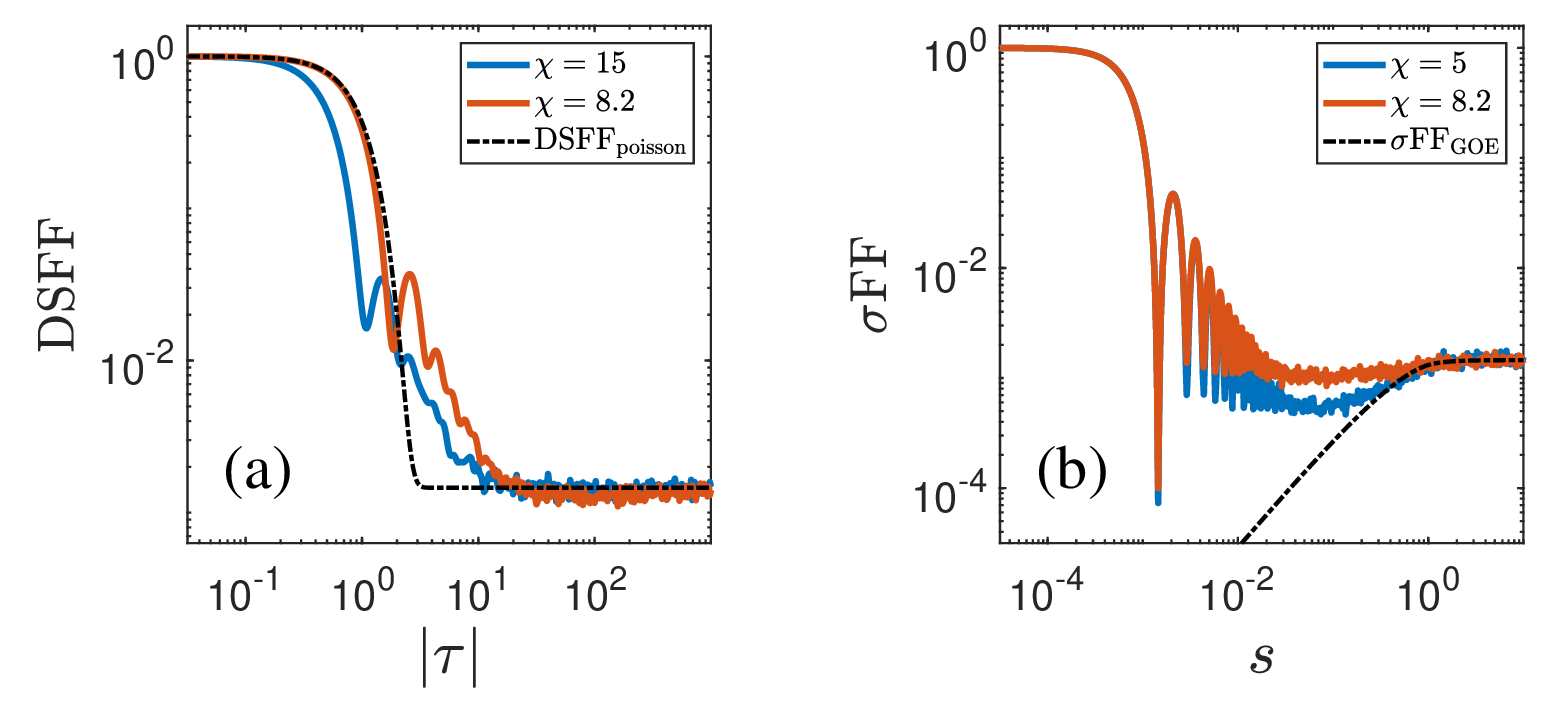}
  \caption{Dynamics of the DSFF and $\sigma$FF for the random-disorder model at $L=14$, averaged over 200 disorder realizations. (a) DSFF for $\chi=8.2$ and $15$; the dashed line indicates the Poisson plateau $1/D$. (b) $\sigma$FF for $\chi=8.2$ and $5$; the dashed line shows the GOE prediction.}
  \label{FIG7}
\end{figure}

We next turn to the TRS-preserving non-Hermitian hard-core-boson chain with random on-site disorder. The Hamiltonian is the same as in Eq.~(\ref{eq:main_hamiltonian}), with the on-site potential term now taken to be a random disorder potential. Specifically, $W_j$ is drawn independently from a uniform distribution, $W_j \in [-\chi,\chi]$, where $\chi$ controls the disorder strength. All other parameters and numerical protocols are chosen identically to those used in the quasiperiodic model, enabling a direct ED--SVD comparison in the TRS-preserving random-disorder setting.

We first characterize spectral correlations using spacing-ratio statistics. From ED, we compute the complex spacing-ratio $\langle r^{\text{E}} \rangle$ from the complex many-body eigenvalues. As shown in Fig.~\ref{FIG5}(a), $\langle r^{\mathrm{E}}\rangle$ evolves from a Ginibre-like value at weak disorder toward the Poisson limit as $\chi$ increases, consistent with a finite-size crossover from an ergodic-like regime to an MBL-like regime. In the SVD framework, the singular-value spacing ratio $\langle r^{\mathrm{S}}\rangle$ in Fig.~\ref{FIG5}(b) exhibits a corresponding drift from a GOE-like value toward Poisson statistics.

To locate the transition, we next analyze state-based diagnostics. Figures~\ref{FIG6}(a) and \ref{FIG6}(b) show the disorder- and bulk-spectrum-averaged half-chain entanglement entropies $\langle S^{\mathrm{E}}\rangle$ and $\langle S^{\mathrm{S}}\rangle$ for three system sizes. With increasing $\chi$, both entropies decrease, reflecting a crossover from volume-law scaling to an area-law-like behavior. The finite-size scaling collapse, however, yields markedly different finite-size transition estimates. In Fig.~\ref{FIG6}(a), an ED-based transition estimate is $\chi_{c}^{\mathrm{E}}\approx 7.7$, whereas Fig.~\ref{FIG6}(b) places the SVD-based transition at a significantly larger value, $\chi_{c}^{\mathrm{S}}\approx 9.5$.

We further test this discrepancy using the IPR. From Figs.~\ref{FIG6}(c) and \ref{FIG6}(d), the scaling collapses yield $\chi_{c}^{\mathrm{E}}\approx 7.7$ from $\langle \mathrm{IPR}^{\mathrm{E}}\rangle$ and $\chi_{c}^{\mathrm{S}}\approx 8.5$ from $\langle \mathrm{IPR}^{\mathrm{S}}\rangle$. Therefore, the ED-based transition points extracted from entanglement and IPR are mutually consistent, while the SVD-based estimates are not only shifted to larger disorder but also show a stronger observable dependence. This systematic ED--SVD mismatch indicates that, in the present TRS-preserving random-disorder non-Hermitian model, SVD-based diagnostics are not reliable for quantitatively determining the MBL transition.

Finally, we test whether the ED--SVD mismatch also appears in SFFs. We choose an intermediate disorder strength $\chi=8.2$, lying between $\chi_{c}^{\mathrm{E}}$ and $\chi_{c}^{\mathrm{S}}$, and compute the DSFF and $\sigma$FF at $L=14$, with the results averaged over $200$ disorder realizations, as shown in Fig.~\ref{FIG7}. In Fig.~\ref{FIG7}(a), the DSFF at $\chi=8.2$ shows no discernible ramp, resembling the strongly disordered case $\chi=15$, which is indicative of Poisson-like spectral correlations in the ED picture. In contrast, Fig.~\ref{FIG7}(b) shows that the $\sigma$FF at the same $\chi=8.2$ still exhibits a weak ramp, and at weaker disorder (e.g., $\chi=5$) it develops a clear dip--ramp--plateau structure, characteristic of an ergodic regime in the SVD picture. Therefore, at the same point $\chi=8.2$, DSFF and $\sigma$FF lead to different conclusions. Together with the shifted SVD finite-size transition estimates in Fig.~\ref{FIG6}, this shows that in TRS-preserving non-Hermitian systems the SVD-based diagnostics can misplace the transition and misidentify the phase, and hence are not reliable for determining the MBL transition in the random-disorder model.

\section{Stark Model}

\begin{figure}[t]
  \includegraphics[width=\linewidth]{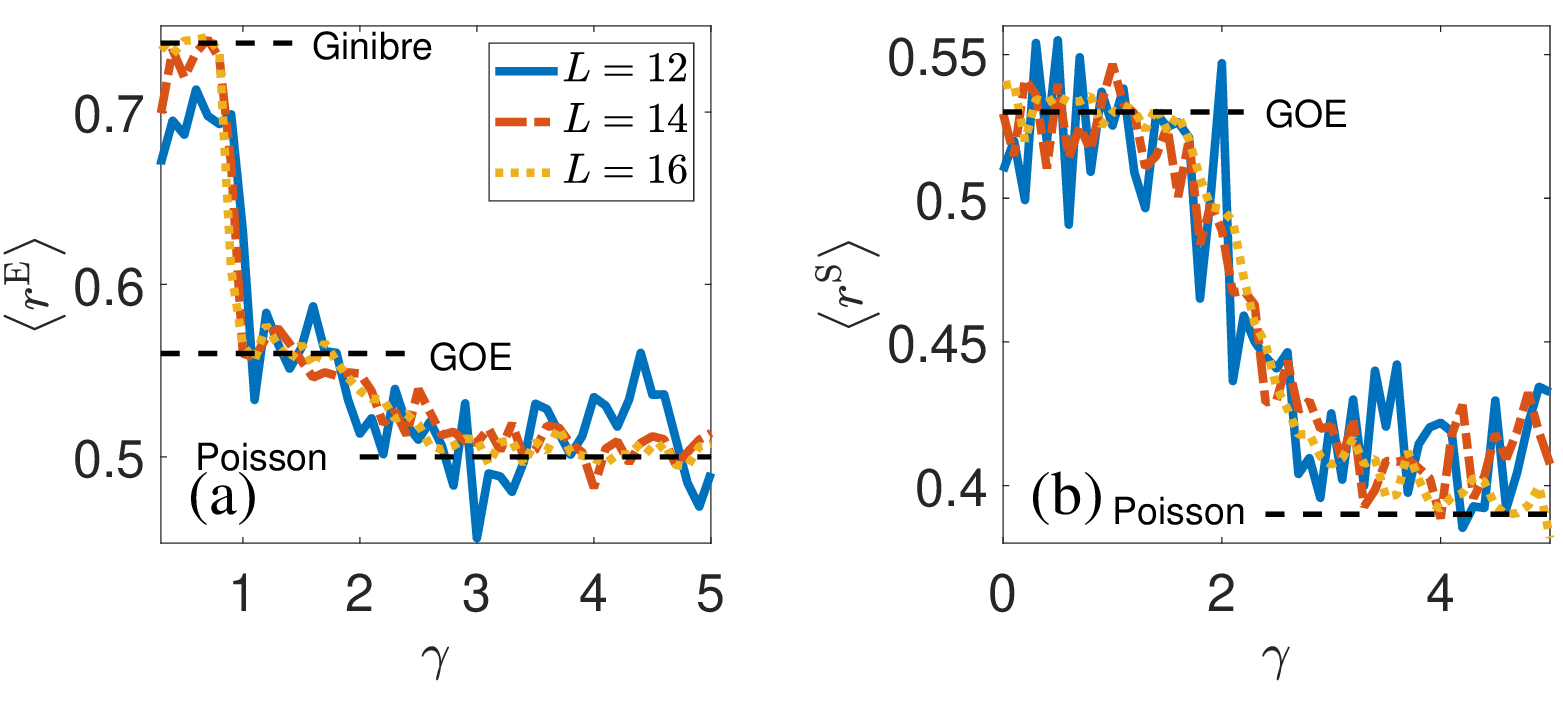}
  \caption{Bulk-spectrum-averaged spacing ratios (a) $\langle r^{\mathrm{E}}\rangle$ and (b) $\langle r^{\mathrm{S}}\rangle$ versus Stark tilt $\gamma$ for different system sizes. Dashed lines indicate the relevant benchmark values for Ginibre-like, GOE-like, and Poisson-like regimes, as applicable.}
  \label{FIG8}
\end{figure}

\begin{figure}[t]
  \includegraphics[width=\linewidth]{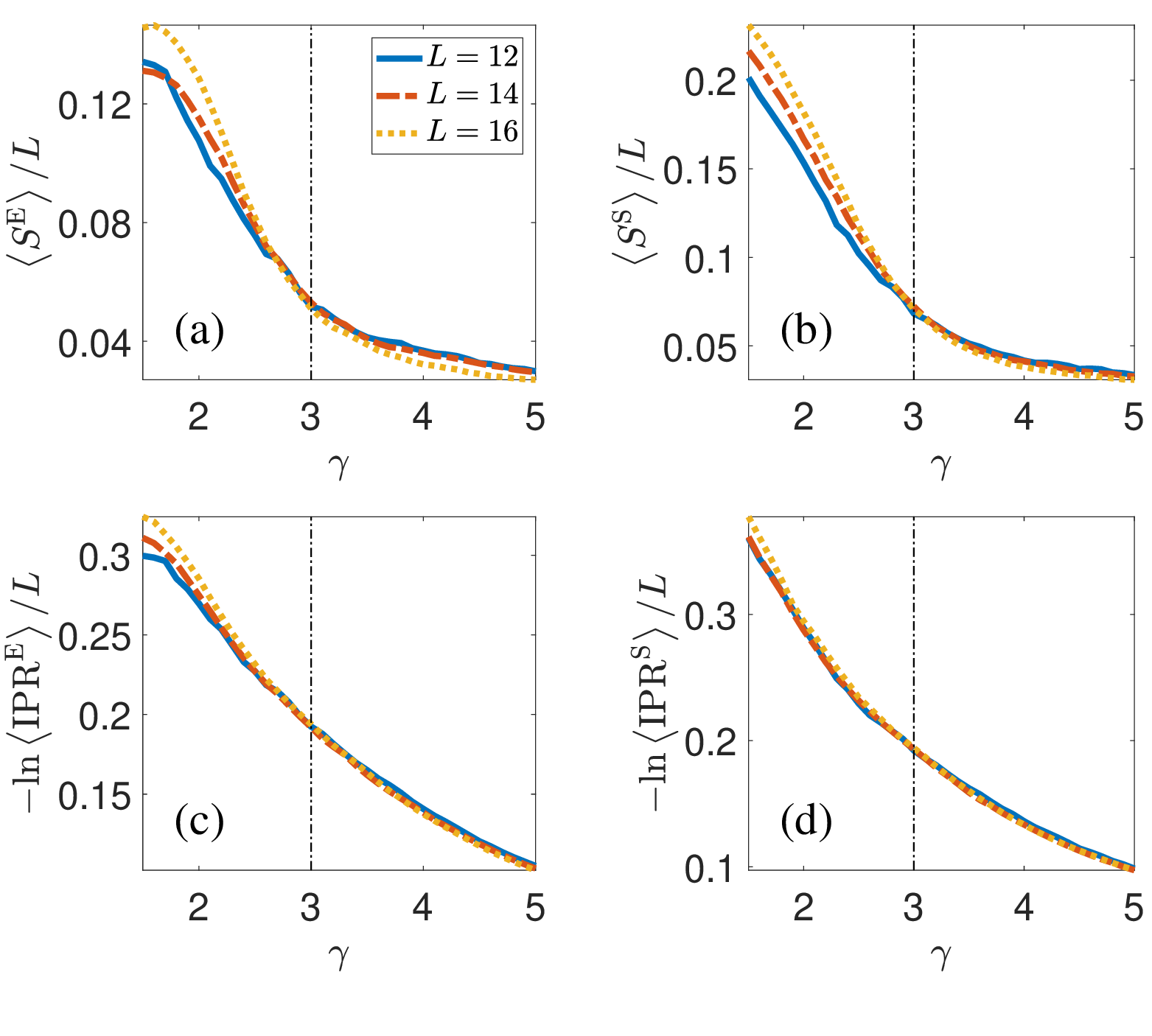}
  \caption{Half-chain entanglement entropy and IPR versus Stark tilt $\gamma$ for different system sizes. Entanglement: (a) $\langle S^{\mathrm{E}}\rangle$ and (b) $\langle S^{\mathrm{S}}\rangle$. IPR: (c) $\langle \mathrm{IPR}^{\mathrm{E}}\rangle$ and (d) $\langle \mathrm{IPR}^{\mathrm{S}}\rangle$. Dashed lines mark the finite-size transition estimates obtained from the scaling analysis.}
  \label{FIG9}
\end{figure}

We now turn to a clean TRS-preserving non-Hermitian hard-core-boson chain in a Stark potential. The on-site potential is $W_{j}=-\gamma (j-1)+\beta\Big(\frac{j-1}{L-1}\Big)^{2}$, where $\gamma$ is the linear tilt strength and $\beta$ sets a weak nonlinear term \cite{SchulzHooleyPRL2019,WannierRMP1962,WannierPR1960}. The nonlinear term is introduced to lift the strong degeneracies of a purely linear Stark ladder and thus restore standard level-statistics behavior. In the following we fix $\beta=0.5$. Since the model is clean, we do not perform disorder averaging.

We first examine spectral statistics. Figure~\ref{FIG8}(a) shows the ED-based complex spacing ratio $\langle r^{\mathrm{E}}\rangle$. At small $\gamma$, $\langle r^{\mathrm{E}}\rangle$ is close to the Ginibre benchmark. As $\gamma$ increases, $\langle r^{\mathrm{E}}\rangle$ crosses over toward a GOE-like value, suggesting that the spectrum gradually becomes real and the level correlations become Hermitian-like. In this regime, where most eigenvalues lie close to the real axis, the complex-spacing analysis effectively approaches the usual real-spectrum spacing-ratio characterization, allowing comparison with the GOE benchmark. Upon further increasing $\gamma$, $\langle r^{\mathrm{E}}\rangle$ drifts toward the Poisson limit, consistent with entering an MBL-like regime. This ED sequence (Ginibre-like $\rightarrow$ GOE-like $\rightarrow$ Poisson-like) is consistent with the conclusion of Ref.~\cite{LiuXuPRB2023}. In the SVD approach, Fig.~\ref{FIG8}(b) shows that the singular-value spacing ratio $\langle r^{\mathrm{S}}\rangle$ starts near the GOE value at small $\gamma$ and crosses over to the Poisson value as $\gamma$ increases, indicating a transition from ergodic to MBL.

To locate the transition point, we compute eigenstate- and singular-vector-based diagnostics. Figure~\ref{FIG9} shows the half-chain entanglement entropies $\langle S^{\mathrm{E}}\rangle$ and $\langle S^{\mathrm{S}}\rangle$ and the IPRs $\langle \mathrm{IPR}^{\mathrm{E}}\rangle$ and $\langle \mathrm{IPR}^{\mathrm{S}}\rangle$. The finite-size scaling collapse yields a consistent finite-size transition estimate, $\gamma_{c}\approx 3$, as extracted from both ED and SVD, and from both entanglement and IPR. Therefore, for this clean Stark model we do not observe the systematic ED--SVD shift found in the previous two TRS-preserving disordered models. We further verify that this ED--SVD agreement in the Stark model persists over a range of nonreciprocity strengths $g$, as shown in Appendix~\ref{app:stark_phase_diagram}.

However, this agreement should still be interpreted with caution. As discussed in Sec.~\ref{sec:discussion} and supported by the bulk-subspace overlap analysis in Appendix~\ref{app:subspace_overlap}, the consistency between ED and SVD in the Stark model appears to arise because the dominant monotonic gradient structure is relatively robust under the construction of $\hat H^\dagger\hat H$, so that the ED and SVD bulk subspaces nearly coincide. This mechanism is not generic: in the quasiperiodic and random-disorder models, the corresponding bulk-subspace overlap is much smaller, and SVD-based diagnostics systematically shift the transition estimates. Therefore, SVD-based diagnostics cannot be regarded as generally reliable quantitative tools for determining MBL transitions in TRS-preserving non-Hermitian systems, and ED-based diagnostics remain necessary.

\section{Discussion: Structural origin of the ED--SVD mismatch}
\label{sec:discussion}

The ED--SVD discrepancy can be understood from the fact that the two approaches probe different operators. ED-based diagnostics are constructed from the eigenvalues and right eigenvectors of $\hat H$, whereas SVD-based diagnostics are determined by the singular values and right singular vectors, i.e., by the auxiliary Hermitian operator $\hat H^\dagger\hat H$. For a non-normal Hamiltonian with $[\hat H,\hat H^\dagger]\neq 0$, the eigenvectors of $\hat H$ and those of $\hat H^\dagger\hat H$ need not span the same local bulk subspace or encode the same spatial and entanglement structure. Therefore, agreement between ED and SVD is not guaranteed even when the Hamiltonian preserves TRS. In this sense, TRS is not the criterion that controls the reliability of SVD; the relevant question is whether $\hat H^\dagger\hat H$ preserves the effective bulk-state structure of $\hat H$. Thus, the role of TRS in the present work is to define the symmetry setting and the associated real-to-complex spectral phenomenology, but it does not remove the structural distinction between $\hat H$ and $\hat H^\dagger\hat H$ that controls the ED--SVD comparison.

Physically, this distinction is particularly important in the quasiperiodic and random-disorder models. In these cases, the construction of $\hat H^\dagger \hat H$ mixes the nonreciprocal hopping, interactions, and spatially varying potentials into an effective Hermitian operator whose local structure is not simply equivalent to that of the original Hamiltonian. As a result, the disorder landscape experienced by the singular vectors can differ from that governing the right eigenstates of $\hat H$, leading to shifted localization crossovers. This explains why, in both the quasiperiodic and random-disorder models, SVD-based entanglement and IPR diagnostics systematically place the transition at larger disorder strength than the ED-based diagnostics.

In the Stark model, by contrast, the dominant spatial structure is a monotonic linear gradient. This leading gradient structure is more robust under the construction of $\hat H^\dagger \hat H$ than a quasiperiodic or random potential landscape. Consequently, the SVD bulk singular-vector subspace remains much more closely aligned with the ED bulk right-eigenstate subspace. This is consistent with the near-unity bulk-subspace overlap and explains why ED and SVD give consistent transition estimates in the Stark model.

This structural distinction is directly supported by the bulk-subspace overlap analysis in Appendix~\ref{app:subspace_overlap}. In the quasiperiodic and random-disorder models, the overlap between the ED bulk eigensubspace and the SVD bulk singular-vector subspace remains far from unity, indicating that the singular vectors encode a different state structure from the ED eigenstates. In the Stark model, by contrast, the overlap is close to unity over the relevant parameter regime, indicating that the two bulk subspaces nearly coincide. Thus, the success or failure of SVD is controlled not by TRS alone, but by whether $\hat H^\dagger\hat H$ preserves the same effective bulk-state structure as $\hat H$.

As an additional check, Appendix~\ref{app:suppressed_nhse} shows that the ED--SVD discrepancy persists in quasiperiodic and random-disorder models with staggered nonreciprocal hopping, where the conventional NHSE associated with a uniform nonreciprocal bias is expected to be suppressed. This indicates that the mismatch is not simply a consequence of a uniform skin effect, but is instead tied to the structural difference between $\hat H$ and $\hat H^\dagger\hat H$.

A further manifestation of the same structural distinction is discussed in Appendix~\ref{app:global_shift}. Under a global shift $\hat H\rightarrow \hat H+c\hat I$, ED-based eigenstate diagnostics remain invariant, whereas SVD-based quantities can change because the auxiliary operator becomes $\hat H_c^\dagger\hat H_c=\hat H^\dagger\hat H+c\hat H^\dagger+c^\ast\hat H+|c|^2\hat I$. This lack of shift invariance further illustrates that SVD probes $\hat H^\dagger\hat H$ rather than the intrinsic right-eigenstate structure of $\hat H$.

\section{Conclusion}

We have benchmarked SVD-based diagnostics against ED in TRS-preserving non-Hermitian hard-core-boson chains using spectral statistics, eigenstate/singular-vector indicators, and SFFs. For the quasiperiodic and random-disorder models, ED-based entanglement entropy and IPR give mutually consistent finite-size transition estimates, whereas the corresponding SVD-based estimates are systematically shifted to larger disorder strengths. This mismatch is also reflected in the SFFs: at the same intermediate parameters, the DSFF already shows Poisson-like behavior, while the $\sigma$FF can still display a residual ramp structure. By contrast, in the clean Stark model, ED and SVD give consistent transition estimates from both entanglement entropy and IPR.

These results show that SVD-based diagnostics can capture qualitative RMT-to-Poisson trends in TRS-preserving non-Hermitian many-body systems, but are not generically reliable for quantitatively locating the MBL transition. The underlying reason is that SVD probes the auxiliary Hermitian operator $\hat H^\dagger\hat H$, rather than the intrinsic right-eigenstate structure of $\hat H$. Thus, TRS alone does not guarantee the validity of SVD diagnostics; the relevant criterion is whether the bulk singular-vector subspace of $\hat H^\dagger\hat H$ remains aligned with the bulk right-eigenstate subspace of $\hat H$. This criterion is not satisfied in the quasiperiodic and random-disorder models, but is approximately satisfied in the Stark model, as supported by the bulk-subspace overlap analysis.

The non-intrinsic character of SVD-based diagnostics is further illustrated by their sensitivity to a global energy shift. As shown in Appendix~\ref{app:global_shift} and Fig.~\ref{FIG16}, a uniform shift of $\hat H$ leaves ED-based eigenstate diagnostics unchanged, but modifies the SVD-based transition estimate because $\hat H^\dagger\hat H$ itself is changed. SVD-based diagnostics should therefore be interpreted as useful qualitative indicators rather than as generally invariant quantitative probes of non-Hermitian many-body phase boundaries.

We emphasize that all quoted transition values are finite-size estimates obtained within the accessible system sizes, rather than thermodynamic-limit critical points. Nevertheless, the relative ED--SVD shift is robust across several observables in the present data. A broader classification of when $\hat H^\dagger\hat H$ preserves the relevant bulk-state structure of $\hat H$ remains an important direction for future work.

\begin{acknowledgments}
  Z. X. is supported by Quantum Science and Technology-National Science and Technology Major Project (Grant No. 2025ZD0300400), the NSFC (Grant No. 12375016), and Beijing National Laboratory for Condensed Matter Physics (No. 2023BNLCMPKF001). Y. Z. is supported by the NSFC (Grant No. 12474492 and No. 12461160324).
\end{acknowledgments}

\section*{Data Availability}
The data that support the findings of this article are not publicly available. The data are available from the authors upon reasonable request.

\appendix

\section{Finite-size scaling}
\label{app:scaling}

In this appendix, we describe the finite-size scaling procedure used to estimate the transition values quoted in the main text. The scaling analysis is illustrated using the half-chain entanglement entropy, and the same protocol is applied to the IPR data quoted in the main text. Since the accessible system sizes are limited to $L=10,12,14$, the resulting transition parameters should be regarded as finite-size transition estimates rather than thermodynamic-limit critical points.

We use a standard one-parameter finite-size scaling form. For a generic control parameter $\lambda$, the scaling variable is defined as
\begin{equation}
  x = (\lambda-\lambda_c)L^{1/\nu},
  \label{eq:scaling_variable}
\end{equation}
where $\lambda_c$ is the estimated transition point and $\nu$ is the correlation-length exponent. Here $\lambda$ denotes $W$, $\chi$, or $\gamma$ for the quasiperiodic, random-disorder, and Stark models, respectively. For the quasiperiodic model shown explicitly below, $\lambda=W$.

For each trial pair $(\lambda_c,\nu)$, every numerical data point $S(\lambda,L)$ is mapped to a point $(x_i,S_i)$, where $x_i=(\lambda_i-\lambda_c)L_i^{1/\nu}$. We then combine all data points from different system sizes into a single dataset and sort them in ascending order of $x_i$. After this sorting, the corresponding entropy values are denoted by $Y_j$. Thus, $Y_j$ is not a new observable; it is simply the half-chain entanglement entropy associated with the $j$th point of the combined dataset after ordering all points by the scaling variable $x$.

To quantify the quality of the data collapse, we use the cost function \cite{PrasadPRB2024,JanaPRB2024,SuntajsPRB2020}
\begin{equation}
  C_Y =
  \frac{\sum_{j=1}^{N_{\mathrm{total}}-1}|Y_{j+1}-Y_j|}
  {\max(Y_j)-\min(Y_j)}
  -1,
  \label{eq:collapse_cost}
\end{equation}
where $N_{\mathrm{total}}$ is the total number of data points collected over all system sizes and control-parameter values. This cost function measures the normalized total variation, or roughness, of the putative collapsed curve. We determine the optimal values of $\lambda_c$ and $\nu$ by scanning the parameter space and minimizing $C_Y$.

\begin{figure}[htbp]
  \centering
  \includegraphics[width=\linewidth]{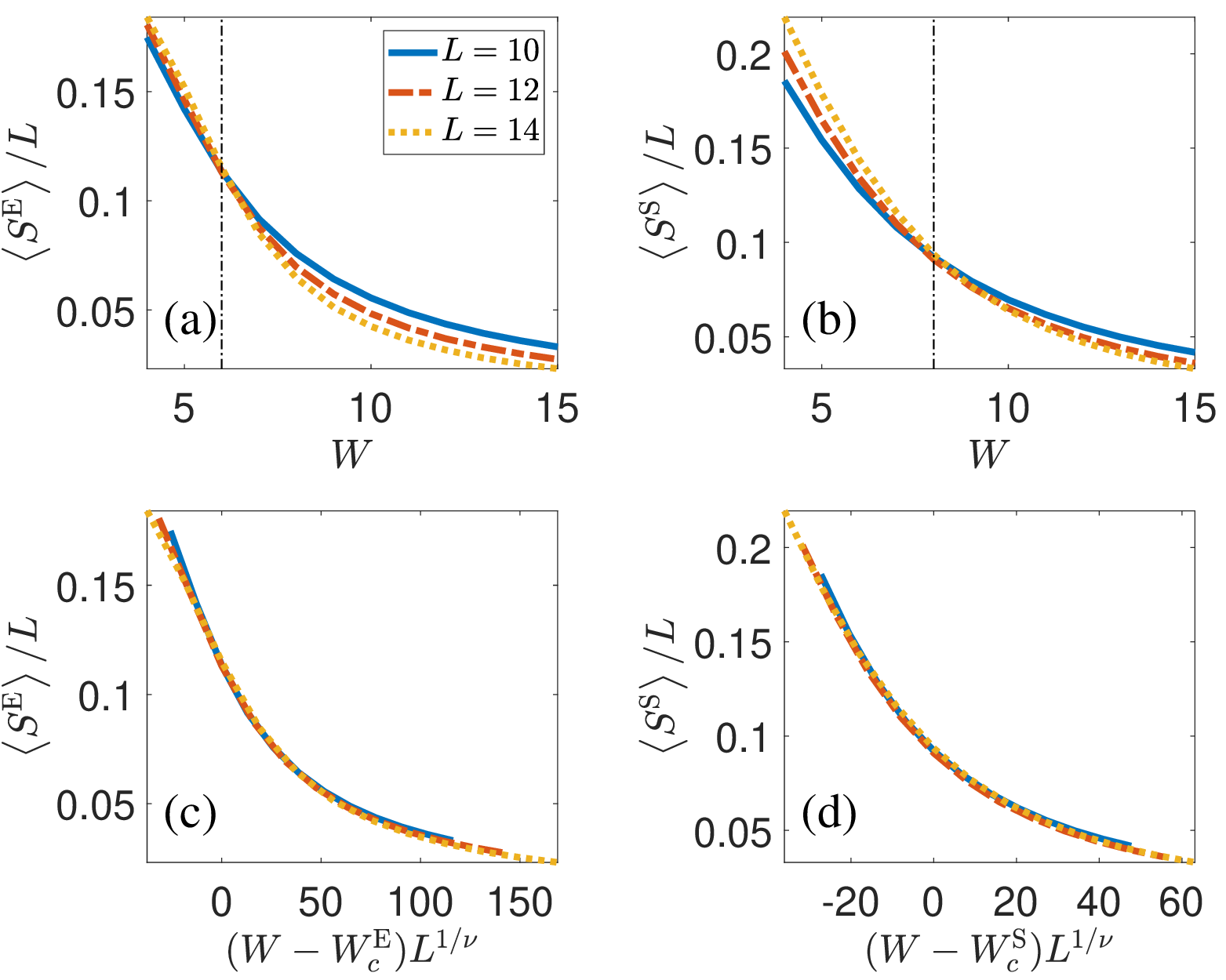}
  \caption{Finite-size scaling analysis of the half-chain entanglement entropy for the quasiperiodic model. (a,b) Entanglement entropy obtained using the ED and SVD methods, respectively, as a function of the quasiperiodic potential strength $W$. (c,d) Corresponding scaling collapses of the data shown in (a) and (b). In (c), the optimized parameters are $W_c^{\mathrm{E}}\approx 6$ and $\nu\approx 0.9$. In (d), the corresponding values are $W_c^{\mathrm{S}}\approx 8$ and $\nu\approx 1.2$.}
  \label{FIG10}
\end{figure}

Figure~\ref{FIG10} shows a representative example of this procedure for the half-chain entanglement entropy in the quasiperiodic model. Panels (a) and (b) show the original ED- and SVD-based entropy data as functions of $W$ for different system sizes. Panels (c) and (d) show the corresponding scaling collapses after plotting the same data as functions of $(W-W_c)L^{1/\nu}$. The optimized parameters are $W_c^{\mathrm{E}}\approx 6$ and $\nu\approx 0.9$ for the ED data, and $W_c^{\mathrm{S}}\approx 8$ and $\nu\approx 1.2$ for the SVD data.

The same scaling protocol is used for the random-disorder model with $\lambda=\chi$ and for the Stark model with $\lambda=\gamma$. To avoid redundancy, Fig.~\ref{FIG10} presents only the representative collapse for the quasiperiodic entanglement data, while the transition estimates obtained from the same procedure for the other models and observables are quoted in the main text.

\section{Angle dependence of the DSFF}
\label{app:dsff_angle}

To further assess the robustness of our results with respect to the choice of the complex time direction, we evaluate the DSFF for additional values of the angle $\theta$, including $\theta = 0$ and $\theta = \pi/5$, for both the quasiperiodic and random models \cite{LiProsenPRL2021, GhoshPRB2022}.

\begin{figure}[htbp]
  \centering
  \includegraphics[width=\linewidth]{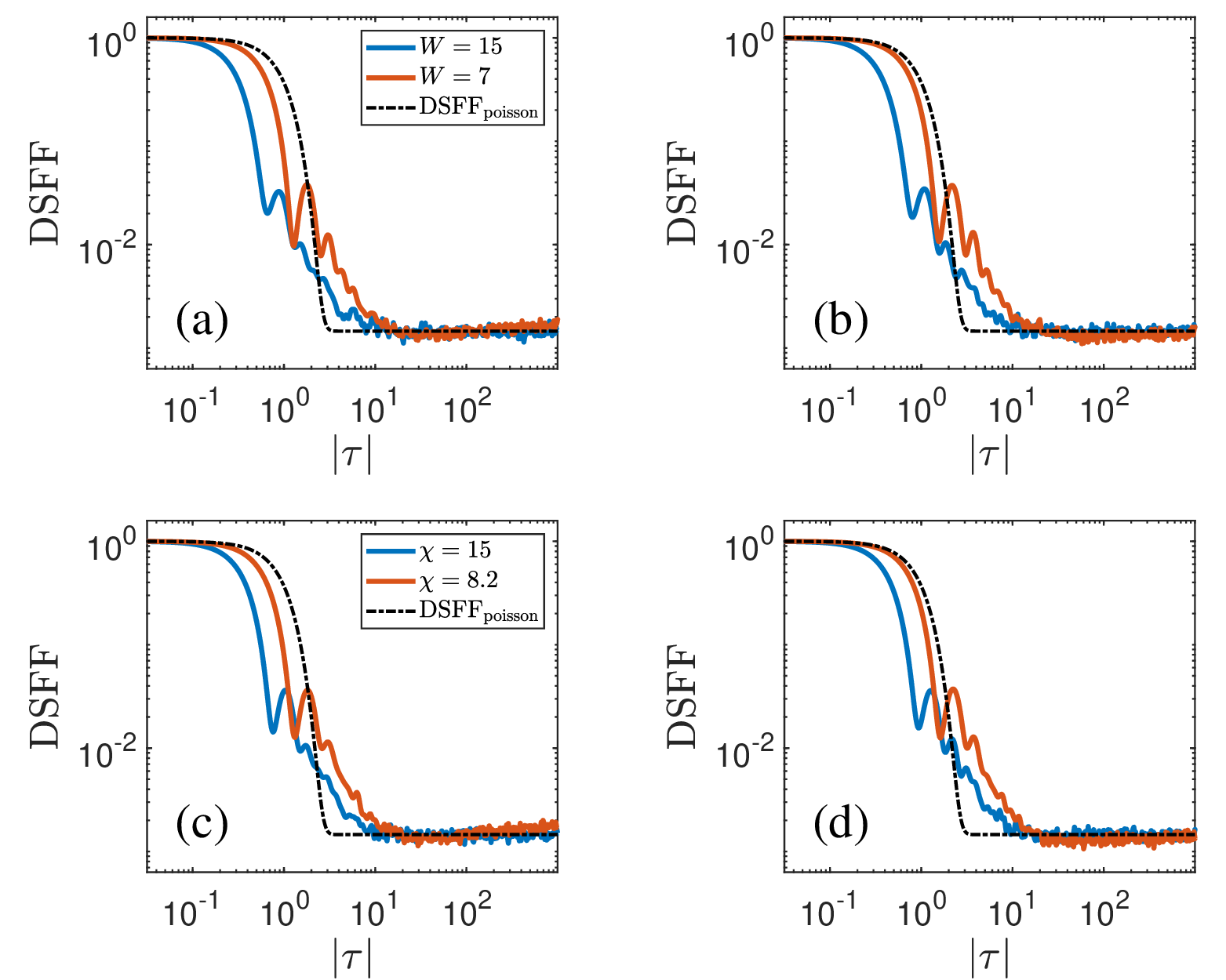}
  \caption{DSFF versus the complex time variable $\tau$ for $L=14$. (a,b) Quasiperiodic model at $W=15$ and $W=7$, respectively, with $\theta=0$ in (a) and $\theta=\pi/5$ in (b). (c,d) Random-disorder model at $\chi=15$ and $\chi=8.2$, respectively, with $\theta=0$ in (c) and $\theta=\pi/5$ in (d).}
  \label{FIG11}
\end{figure}

As shown in Fig.~\ref{FIG11}, the DSFF exhibits qualitatively similar behavior for all values of $\theta$ considered. In particular, the absence of a well-defined ramp structure persists across these different angles. This behavior contrasts with the $\sigma$FF results shown in the main text, where a ramp structure remains visible in the corresponding parameter regime. Specifically, Figs.~\ref{FIG11}(a) and \ref{FIG11}(b) show the DSFF of the quasiperiodic model as a function of time for disorder strengths $W=15$ and $W=7$, respectively, with $\theta=0$ in Fig.~\ref{FIG11}(a) and $\theta=\pi/5$ in Fig.~\ref{FIG11}(b). Similarly, Figs.~\ref{FIG11}(c) and \ref{FIG11}(d) present the corresponding results for the random-disorder model at $\chi=15$ and $\chi=8.2$, respectively, where Fig.~\ref{FIG11}(c) corresponds to $\theta=0$ and Fig.~\ref{FIG11}(d) to $\theta=\pi/5$.

We thus conclude that the absence of the ramp in the DSFF is not an artifact of the specific choice $\theta = \pi/4$, but rather a robust feature of the DSFF itself. This further supports the qualitative distinction between the DSFF and the $\sigma$FF discussed in the main text.

\section{Phase diagram of the Stark model}
\label{app:stark_phase_diagram}

To further examine the dependence of the MBL transition on the nonreciprocity parameter $g$, we extend our analysis of the Stark model beyond the fixed value $g=0.5$ considered in the main text \cite{LiuXuPRB2023,LiYuPRA2023}.

\begin{figure}[htbp]
  \centering
  \includegraphics[width=\linewidth]{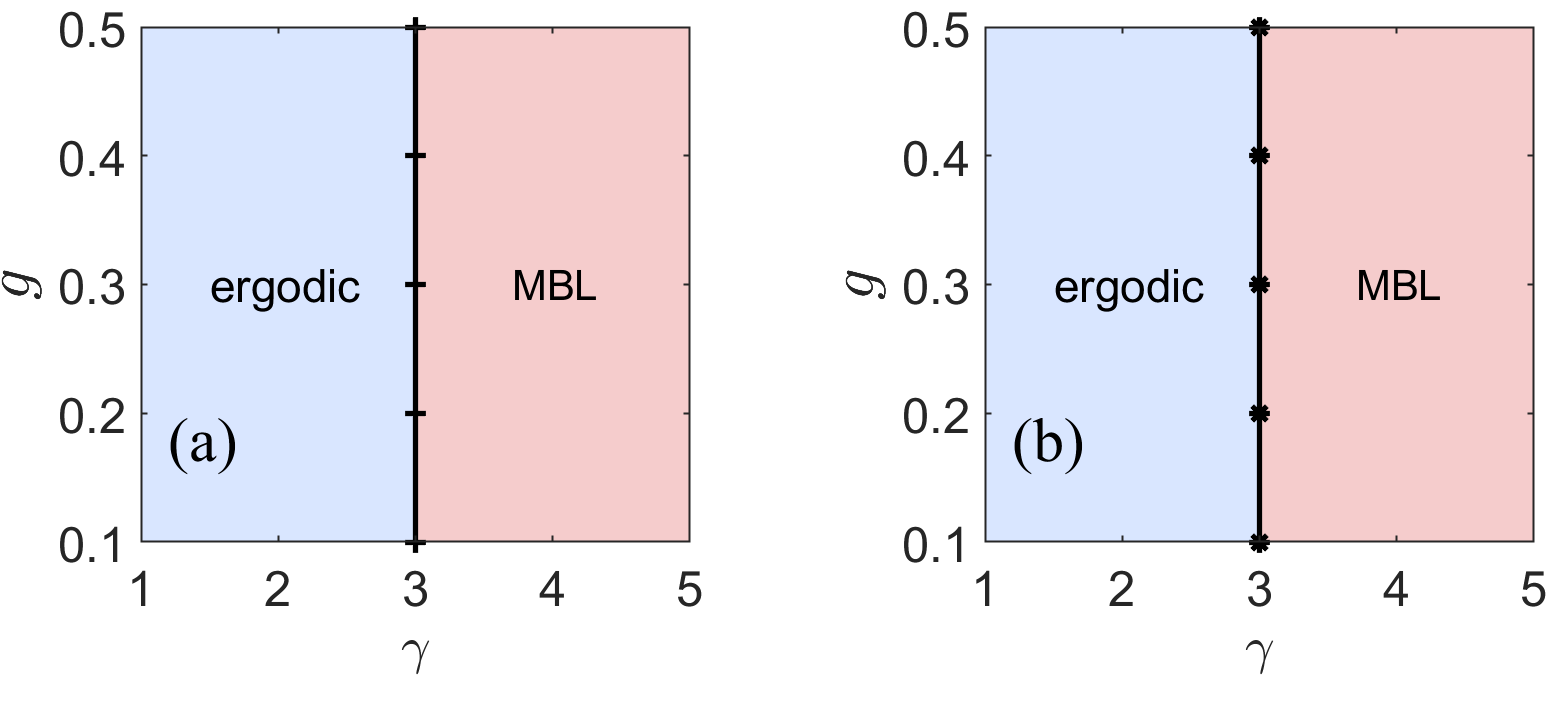}
  \caption{Phase diagram of the Stark model in the $(g, \gamma)$ plane. The horizontal axis denotes the linear tilt strength $\gamma$, while the vertical axis represents the non-Hermitian strength $g$. (a) Phase boundary obtained using ED-based diagnostics. (b) Phase boundary obtained using SVD-based diagnostics.}
  \label{FIG12}
\end{figure}

Specifically, we compute the finite-size transition estimate $\gamma_c$ over a range of $g$ values using both ED- and SVD-based diagnostics. The resulting phase diagram in the $(g,\gamma)$ plane is shown in Fig.~\ref{FIG12}, where Fig.~\ref{FIG12}(a) presents the results obtained from the ED-based diagnostics, while Fig.~\ref{FIG12}(b) shows the corresponding results from the SVD-based diagnostics.

We find that the transition lines extracted from ED and SVD remain in good agreement across the range of $g$ considered. In particular, the two approaches yield nearly identical $\gamma_c$ within numerical resolution, indicating that the consistency observed at $g=0.5$ persists within the finite-size and parameter regime accessible to our numerics.

These results demonstrate that, within the finite-size and parameter regime considered here, the agreement between ED- and SVD-based diagnostics in the Stark model is robust with respect to variations in the nonreciprocity strength.

\section{Angular statistics of the complex spacing ratio}
\label{app:angular_statistics}

\begin{figure}[htbp]
  \centering
  \includegraphics[width=\linewidth]{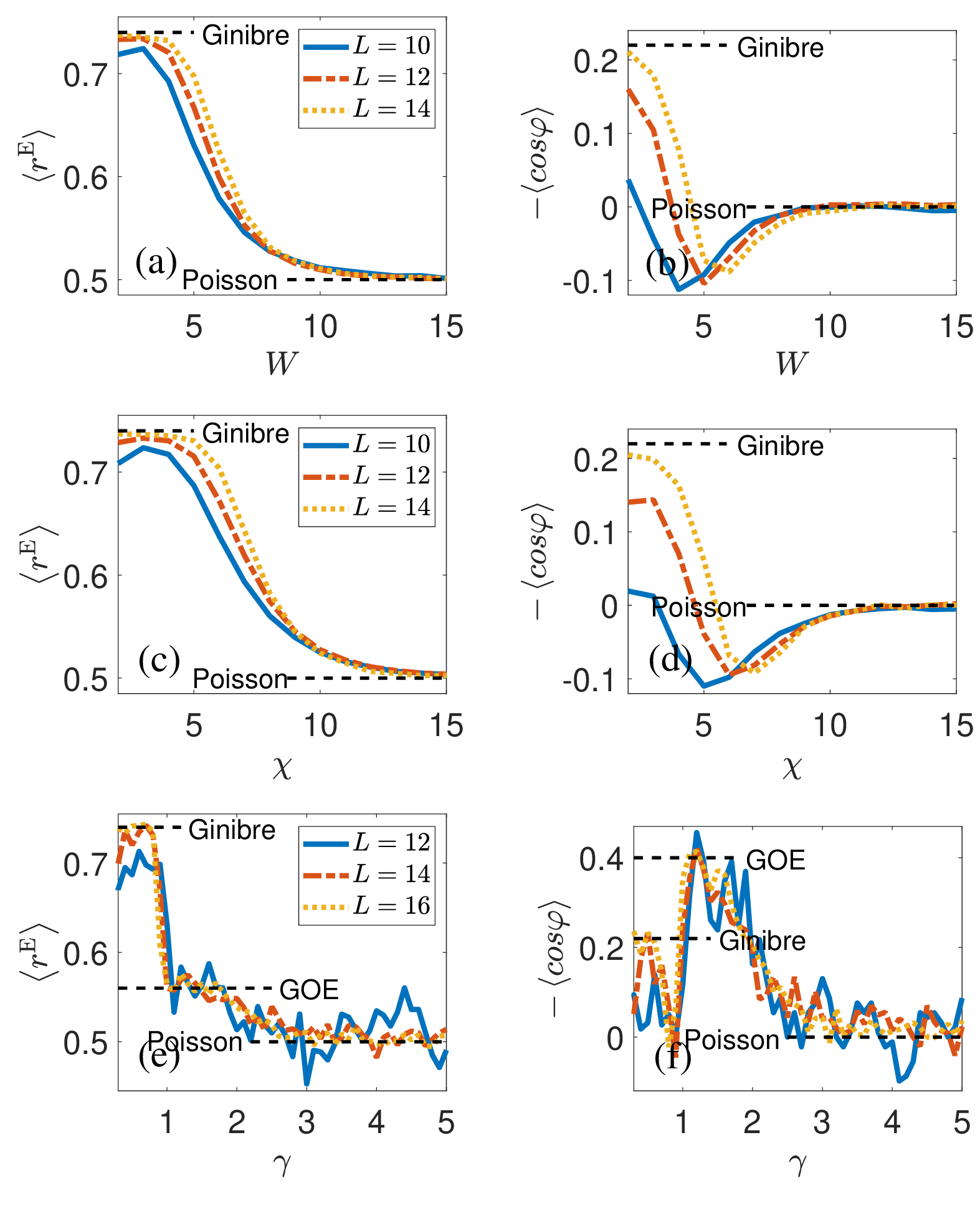}
  \caption{Complex spacing ratio statistics for different models. (a,b) Radial and angular statistics of the complex spacing ratio for the quasiperiodic model as functions of the quasiperiodic disorder strength $W$. (c,d) Radial and angular statistics for the random-disorder model as functions of the disorder strength $\chi$. (e,f) Radial and angular statistics for the Stark model as functions of the linear tilt strength $\gamma$. In each case, the horizontal axis denotes the corresponding control parameter, while the vertical axis represents the radial (angular) component in panels (a,c,e) [panels (b,d,f)].}
  \label{FIG13}
\end{figure}

To further complement the radial analysis of the complex spacing ratio presented in the main text, we examine here the angular statistics of the complex spacing ratio defined in Eq.~(\ref{eq:complex_spacing_ratio}) for the quasiperiodic, random-disorder, and Stark models. We characterize the angular statistics through the quantity $-\langle \cos \varphi \rangle$, where $\varphi$ denotes the angular part of the complex spacing ratio in the complex-energy plane. The corresponding results are shown in Fig.~\ref{FIG13}. Specifically, Figs.~\ref{FIG13}(a) and \ref{FIG13}(b) present the radial and angular statistics for the quasiperiodic model, respectively. Figs.~\ref{FIG13}(c) and \ref{FIG13}(d) show the corresponding results for the random-disorder model, while Figs.~\ref{FIG13}(e) and \ref{FIG13}(f) display those for the Stark model.

For the quasiperiodic and random-disorder models, the angular statistics exhibit qualitatively similar behavior. In the weak-disorder regime, the angular distribution is consistent with the Ginibre ensemble, characterized by $-\langle \cos\varphi \rangle \approx 0.22$ \cite{SaPRX2020,SutharWangPRB2022,GhoshPRB2022}.
As the disorder strength increases, the system gradually approaches the Poisson limit, $-\langle \cos\varphi \rangle \approx 0$,	which is consistent with localized spectra and the absence of level repulsion.

Interestingly, in the intermediate regime near the transition region, we observe negative values of $-\langle \cos\varphi \rangle$. Such behavior deviates from both the Ginibre and Poisson limits and may reflect finite-size spectral anisotropies or the coexistence of different local spectral structures near the crossover regime. A detailed characterization of this angular anomaly is beyond the scope of the present work.

For the Stark model, the angular statistics display a different evolution compared with the quasiperiodic and random-disorder models. In the weak-tilt regime, the results are consistent with Ginibre statistics. As the Stark linear tilt strength increases, both the radial and angular statistics evolve from the Ginibre regime to an intermediate GOE regime before eventually approaching the Poisson limit at large tilt strength. The consistency between the radial and angular statistics indicates that the Stark model undergoes a clear crossover from non-Hermitian Ginibre statistics to Hermitian-like GOE statistics prior to localization. This intermediate GOE regime may originate from the increasing tendency of the spectrum to become effectively real in the Stark model, so that the TRS-preserving non-Hermitian spectrum crosses over to Hermitian-like GOE statistics.

Overall, the angular statistics provide complementary information about the spectral structure in the complex plane, especially regarding spectral anisotropy and the real-to-complex evolution. However, they do not qualitatively alter the conclusions of the main text, nor do they resolve the discrepancies observed between the ED- and SVD-based diagnostics. This is because the ED--SVD discrepancy primarily originates from the structural difference between the right-eigenstate subspace of $\hat H$ and the singular-vector subspace of $\hat H^\dagger\hat H$, whereas the angular statistics characterize only the complex ED spectrum and have no direct counterpart in the nonnegative singular-value spectrum.

\section{TRS-preserving models with staggered nonreciprocal hopping}
\label{app:suppressed_nhse}

To further clarify the role of time-reversal symmetry and to check that the observed ED--SVD discrepancy is not tied to a uniform nonreciprocal bias, we additionally consider TRS-preserving interacting chains with staggered nonreciprocal hopping under periodic boundary conditions. The staggered form of the nonreciprocity cancels the net nonreciprocal bias over a two-site unit cell and is therefore expected to suppress the conventional non-Hermitian skin effect (NHSE) associated with a uniform imaginary gauge flux under open-boundary conditions. We examine three representative on-site potentials: quasiperiodic, random-disorder, and Stark potentials. The Hamiltonian is written as $\hat{H}^{\prime}=\hat{H}^{\prime}_I+\hat{W}^{\prime}$, with
\begin{equation}
  \hat{H}^{\prime}_{I}
  =\sum_{j}\left[-J\left(
    t^{(R)}_j \, \hat b^\dagger_{j+1} \hat b_j+
    t^{(L)}_j \, \hat b^\dagger_j \hat b_{j+1}
    \right) +U\hat{n}_{j}\hat{n}_{j+1}\right],
  \label{eq:staggered_hamiltonian}
\end{equation}
and $\hat{W}^{\prime}=\sum_j W_j^{\prime}\hat{n}_j$ denotes the on-site potential term. The nonreciprocal hopping amplitudes take the staggered form
\begin{equation}
  t^{(R)}_j = e^{(-1)^j g}, \qquad
  t^{(L)}_j = e^{-(-1)^j g},
\end{equation}
with $g=0.5$. The hopping amplitude $J=1$, and the nearest-neighbor interaction strength $U=2$. We focus on the half-filling case, corresponding to a particle density $N/L=1/2$. All hopping amplitudes are real, and thus the Hamiltonian preserves time-reversal symmetry. The staggered asymmetric hopping further removes the net nonreciprocal bias over a two-site unit cell, providing a complementary setting to test the robustness of the ED--SVD discrepancy.
 
\begin{figure}[htbp]
  \centering
  \includegraphics[width=\linewidth]{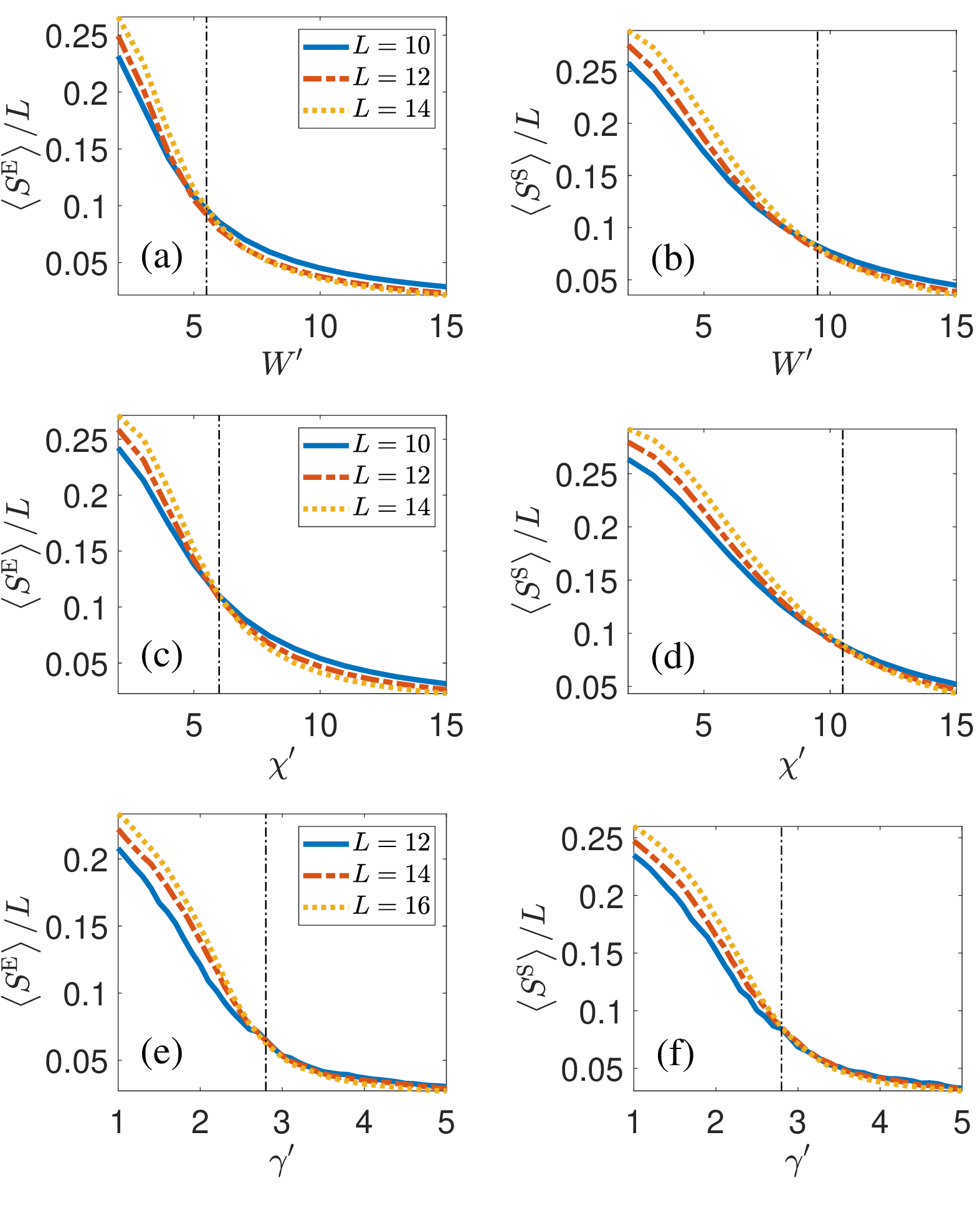}
  \caption{Half-chain entanglement entropy for TRS-preserving non-Hermitian models with staggered nonreciprocal hopping, where the conventional NHSE associated with a uniform nonreciprocal bias is expected to be suppressed. Panels (a) and (b) show the quasiperiodic model as a function of the quasiperiodic potential strength $W^\prime$, obtained using ED- and SVD-based diagnostics, respectively. The corresponding transition estimates are $W_c^{\prime\mathrm{E}}\approx 5.5$ and $W_c^{\prime\mathrm{S}}\approx 9.5$. Panels (c) and (d) show the random-disorder model, yielding $\chi_c^{\prime\mathrm{E}}\approx 6$ and $\chi_c^{\prime\mathrm{S}}\approx 10.5$. Panels (e) and (f) show the Stark model, where the ED- and SVD-based diagnostics give a consistent transition estimate $\gamma_c^\prime\approx 2.8$.}
  \label{FIG14}
\end{figure}

For the first model, we consider a quasiperiodic on-site potential of the form $W_j^{\prime}=W^{\prime}\cos(2\pi \alpha j+\phi)$, where $\alpha=(\sqrt{5}-1)/2$ is an irrational number and $\phi$ is a phase offset uniformly sampled within $[0,2\pi)$. The parameter $W^{\prime}$ controls the quasiperiodic potential strength. The half-chain entanglement entropies obtained from ED and SVD are shown in Figs.~\ref{FIG14}(a) and \ref{FIG14}(b). The two approaches again produce substantially different finite-size estimates of the MBL transition. In Fig.~\ref{FIG14}(a), the ED-based analysis gives $W_c^{\prime\mathrm{E}}\approx 5.5$, whereas the SVD-based result in Fig.~\ref{FIG14}(b) yields a significantly larger transition estimate, $W_c^{\prime\mathrm{S}}\approx 9.5$. This discrepancy further demonstrates that, even in a TRS-preserving staggered nonreciprocal setting with suppressed uniform nonreciprocal bias, the SVD-based diagnostics may still deviate considerably from the ED results in quasiperiodic systems.
 
For the second model, the on-site potential is random, where each $W_j^{\prime}$ is independently sampled from a uniform distribution within the interval $[-\chi^{\prime},\chi^{\prime}]$, with $\chi^{\prime}$ controlling the disorder strength. The half-chain entanglement entropies obtained from ED and SVD are shown in Figs.~\ref{FIG14}(c) and \ref{FIG14}(d). The two approaches again produce substantially different finite-size estimates of the MBL transition. In Fig.~\ref{FIG14}(c), the ED-based analysis gives $\chi_c^{\prime\mathrm{E}}\approx 6$, whereas the SVD-based result in Fig.~\ref{FIG14}(d) yields a significantly larger transition estimate, $\chi_c^{\prime\mathrm{S}}\approx 10.5$. This confirms that the ED--SVD mismatch persists also in the random-disorder case in a staggered nonreciprocal setting where the conventional NHSE associated with a uniform nonreciprocal bias is expected to be suppressed.

For the third model, we consider a Stark-type on-site potential, $W_{j}^{\prime}=-\gamma^{\prime} (j-1)+\beta\Big(\frac{j-1}{L-1}\Big)^{2}$, where $\gamma^{\prime}$ denotes the linear tilt strength and $\beta$ controls a weak nonlinear correction. We fix $\beta=0.5$. Figs.~\ref{FIG14}(e) and \ref{FIG14}(f) show the half-chain entanglement entropies $\langle S^{\mathrm{E}}\rangle$ and $\langle S^{\mathrm{S}}\rangle$. In this case, finite-size scaling analysis yields a consistent finite-size transition estimate, $\gamma_c^{\prime}\approx 2.8$, from both ED and SVD. Therefore, the SVD-based diagnostics remain consistent with the ED-based diagnostics for this TRS-preserving Stark model.

These results suggest that the effectiveness of SVD-based diagnostics in TRS-preserving non-Hermitian systems remains strongly model dependent, even in a staggered nonreciprocal setting where the conventional NHSE associated with a uniform nonreciprocal bias is expected to be strongly suppressed.

\section{Bulk-subspace overlap between ED and SVD states}
\label{app:subspace_overlap}

To further quantify the relation between the ED bulk states and the SVD bulk states, we calculate the overlap between the corresponding bulk subspaces. For the ED approach, we construct the bulk right-eigenstate matrix
\begin{equation*}
\Psi_E=\left( |\psi_{i_1}\rangle,\dots,|\psi_{i_{D_b}}\rangle \right),
\end{equation*}
where the selected ED states are those closest to the center of the complex spectrum according to the distance in the complex-energy plane. The SVD states are selected from the central part of the ordered singular-value spectrum, using the same number of states $D_b$. For the SVD approach, the bulk right-singular-vector matrix is defined as
\begin{equation*}
V_S=\left( |v_{j_1}\rangle,\dots,|v_{j_{D_b}}\rangle \right).
\end{equation*}
Since the right eigenstates of a non-Hermitian Hamiltonian are generally nonorthogonal, we first orthogonalize the two subspaces through QR decomposition
\begin{equation*}
\Psi_E = Q_E R_E, \qquad V_S = Q_S R_S,
\end{equation*}
with
\begin{equation*}
Q_E^\dagger Q_E = I, \qquad Q_S^\dagger Q_S = I.
\end{equation*}

The overlap between the two bulk subspaces is then characterized by

\begin{equation}
\mathcal O=\frac{1}{D_b}\mathrm{Tr}(P_E P_S)
=\frac{1}{D_b}\left\|Q_E^\dagger Q_S\right\|_F^2,
  \label{eq:overlap}
\end{equation}
where
\begin{equation*}
P_E=Q_E Q_E^\dagger,\qquad
P_S=Q_S Q_S^\dagger,
\end{equation*}
and $\|\cdots\|_F$ denotes the Frobenius norm. The quantity $\mathcal O$ satisfies $0\le \mathcal O\le 1$, where $\mathcal O=1$ indicates that the two bulk subspaces are identical.

\begin{figure}[htbp]
	\centering	
	\includegraphics[width=\linewidth]{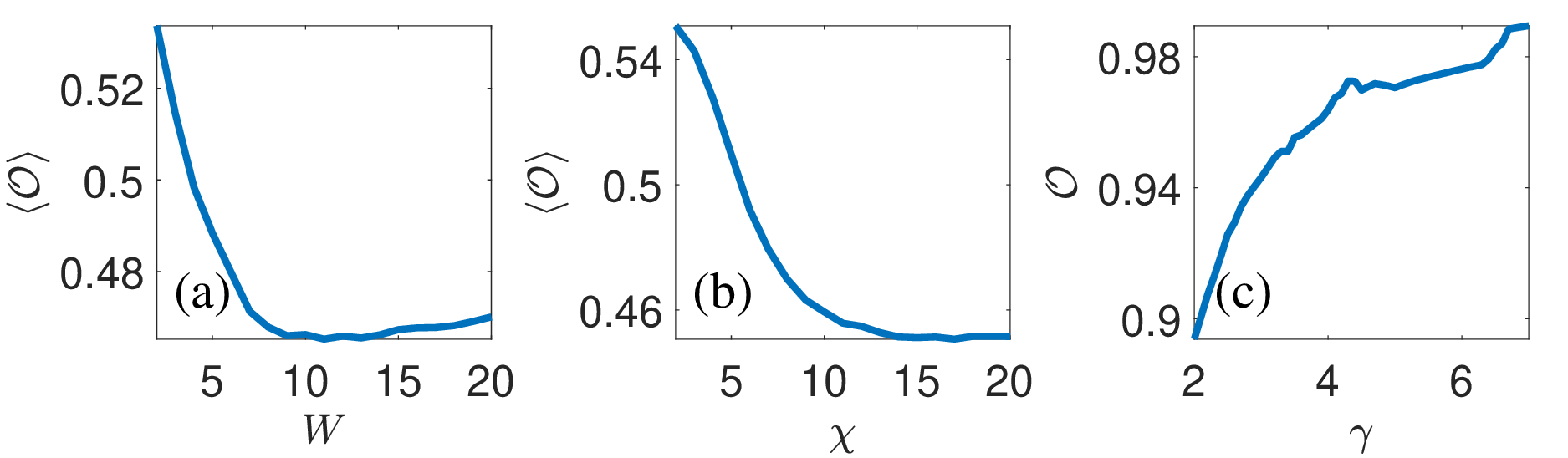}
	\caption{Overlap $\mathcal O$ between the ED bulk subspace and the SVD bulk subspace for (a) the quasiperiodic model, (b) the random-disorder model, and (c) the Stark model. Panels (a) and (b) are obtained for $L=12$ with averaging over $200$ disorder realizations, while panel (c) is calculated for $L=14$ without disorder averaging.}
	\label{FIG15}
\end{figure}

The overlap $\mathcal O$ between the ED bulk subspace and the SVD bulk subspace for the quasiperiodic, random-disorder, and Stark models is shown in Figs.~\ref{FIG15}(a)--\ref{FIG15}(c), respectively. For the quasiperiodic and random-disorder models, the system size is fixed at $L=12$, and the results are averaged over $200$ disorder realizations. In contrast, the Stark model is calculated at $L=14$ without disorder averaging. The horizontal axes correspond to the quasiperiodic strength $W$, the random disorder strength $\chi$, and the Stark linear tilt strength $\gamma$, while the vertical axis represents the overlap $\mathcal O$ (or the disorder-averaged overlap $\langle \mathcal O \rangle$ for the quasiperiodic and random-disorder cases).

One can observe that, for both the quasiperiodic and random-disorder models, the overlap remains substantially below unity, indicating that the SVD bulk subspace does not faithfully reproduce the ED bulk eigensubspace. In contrast, for the Stark model, the overlap approaches $\mathcal O\simeq 1$, demonstrating that the SVD bulk states and the ED bulk states span nearly the same subspace. These results provide direct evidence that the effectiveness of the SVD-based characterization is strongly model dependent. In particular, the SVD bulk subspace can faithfully reproduce the ED bulk structure in the Stark system, whereas in the quasiperiodic and random-disorder systems it deviates significantly from the true ED bulk eigensubspace.

\section{Effect of a global energy shift on SVD diagnostics}
\label{app:global_shift}

We finally comment on the effect of a global energy shift, which further illustrates the structural difference between ED- and SVD-based diagnostics. Under the transformation
\begin{equation}
  \hat{H}\rightarrow \hat{H}_c=\hat{H}+c\hat{I},
\end{equation}
with a complex constant $c$, the right eigenstates of $\hat{H}$ are unchanged and all eigenvalues are shifted uniformly, $E_n\rightarrow E_n+c$. Consequently, ED-based eigenstate diagnostics such as entanglement entropy and IPR are invariant, and spacing-ratio statistics based on eigenvalue differences are also unchanged.

By contrast, the SVD of $\hat{H}_c$ is governed by
\begin{equation}
  \hat{H}_c^{\dagger}\hat{H}_c
  =
  \hat{H}^{\dagger}\hat{H}
  +c\hat{H}^{\dagger}
  +c^{\ast}\hat{H}
  +|c|^2\hat{I}.
  \label{eq:shift_hdagh}
\end{equation}
Except in special cases where the additional terms do not modify the eigenvectors of $\hat{H}^{\dagger}\hat{H}$, the singular values and singular vectors are therefore not invariant under a global shift of the Hamiltonian. This lack of shift invariance is absent in ED-based diagnostics and highlights that SVD probes the auxiliary Hermitian operator $\hat{H}^{\dagger}\hat{H}$ rather than the intrinsic eigenstate structure of $\hat{H}$ itself.

For the main-text results, we fix the Hamiltonian convention specified in Eq.~(\ref{eq:main_hamiltonian}) and do not introduce additional global shifts. Nevertheless, Eq.~(\ref{eq:shift_hdagh}) shows that SVD-based transition estimates can in principle depend on such a convention. This provides another reason why SVD-based diagnostics should be interpreted as useful qualitative indicators rather than as generally invariant quantitative probes of non-Hermitian MBL transitions.

\begin{figure}[htbp]
	\centering	
	\includegraphics[width=\linewidth]{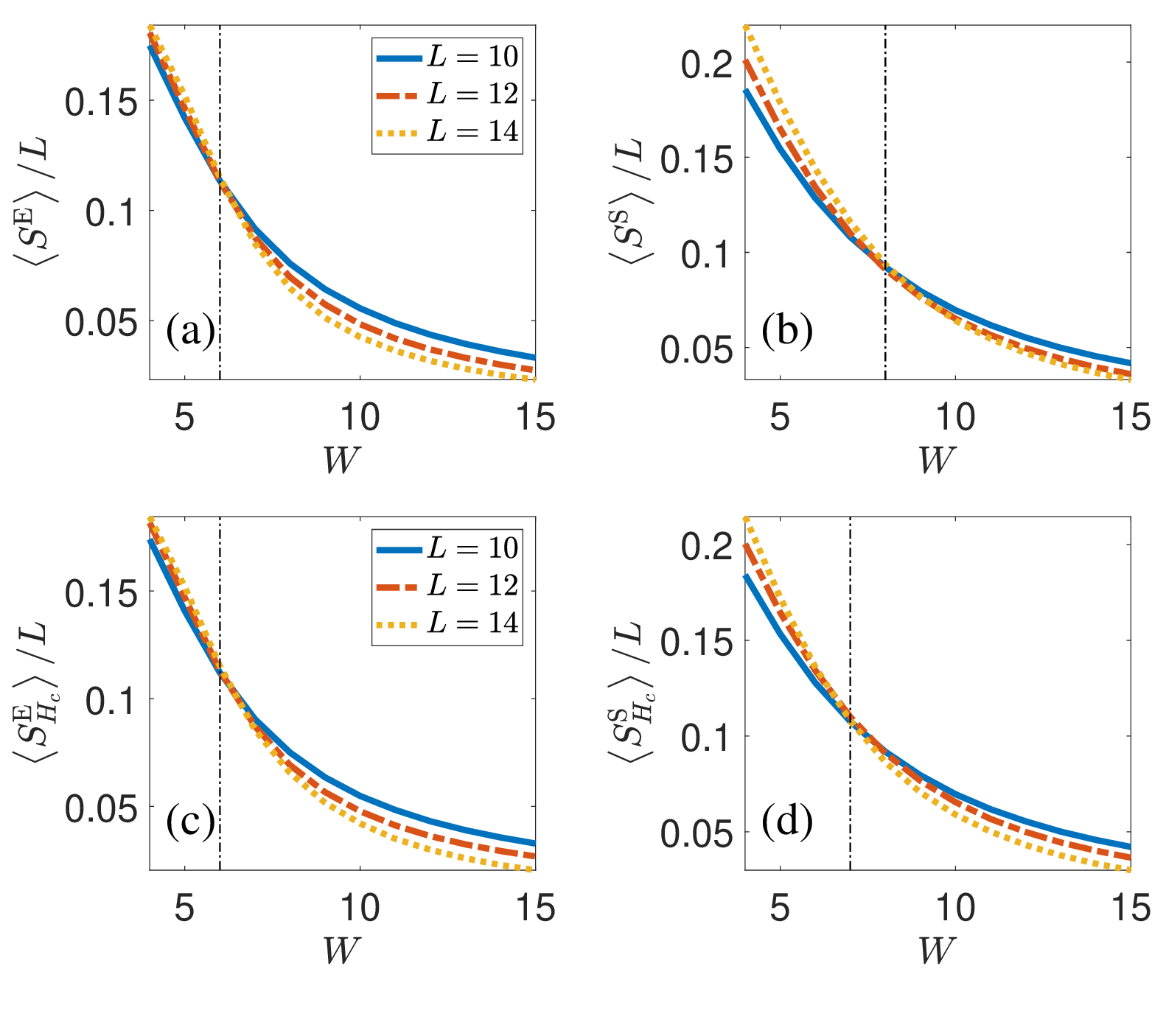}
	\caption{Half-chain entanglement entropy of the quasiperiodic model as a function of the quasiperiodic strength $W$ obtained from the ED and SVD approaches in panels (a) and (b), respectively. Panels (c) and (d) show the corresponding results after introducing a real global spectral shift $c=0.5$ in units of $J$.}
	\label{FIG16}
\end{figure}

To explicitly demonstrate the different responses of ED- and SVD-based diagnostics to a global spectral shift, we further compare the half-chain entanglement entropy obtained before and after introducing a real uniform shift $c=0.5$ in units of $J$. The corresponding results for the quasiperiodic model are shown in Fig.~\ref{FIG16}. Figures~\ref{FIG16}(a) and \ref{FIG16}(b) present the ED- and SVD-based results for the original Hamiltonian, respectively, while Figs.~\ref{FIG16}(c) and \ref{FIG16}(d) show the corresponding results after the spectral shift is introduced.

One can clearly observe that the ED results remain essentially unchanged after the shift, and the crossing point identifying the finite-size transition is preserved at nearly the same quasiperiodic strength, as demonstrated by the close agreement between Figs.~\ref{FIG16}(a) and \ref{FIG16}(c). This behavior is expected because ED-based quantities depend only on the eigenstates of the Hamiltonian and are therefore invariant under a uniform energy shift. In contrast, the SVD-based results exhibit visible modifications after the shift, as shown by the differences between Figs.~\ref{FIG16}(b) and \ref{FIG16}(d). In particular, the estimated transition point $W_{c}^{\mathrm{S}}$ shifts from approximately $8$ to $7$, indicating that the SVD-based transition point is sensitive to the choice of the global spectral convention. These numerical results are fully consistent with the analysis of Eq.~(\ref{eq:shift_hdagh}) and further demonstrate that SVD probes the auxiliary Hermitian operator $\hat{H}^{\dagger}\hat{H}$ rather than the intrinsic eigenstate and spectral structure of the original non-Hermitian Hamiltonian.

\par
\par


\begin{thebibliography}{99}
  \par
  \par
  \bibitem{AbaninRMP2019}{D. A. Abanin, E. Altman, I. Bloch, and M. Serbyn, Colloquium: Many-body localization, thermalization, and entanglement, Rev. Mod. Phys. \textbf{91}, 021001 (2019).}
  \bibitem{HuseP2016}{D. A. Huse, Many-body localization needs a bath, Physics \textbf{9}, 76 (2016).}
  \bibitem{OganesyanPRB2007}{V. Oganesyan and D. A. Huse, Localization of interacting fermions at high temperature, Phys. Rev. B \textbf{75}, 155111 (2007).}
  \bibitem{NandkishoreARCMP2015}{R. Nandkishore and D. A. Huse, Many-body localization and thermalization in quantum statistical mechanics, Annu. Rev. Condens. Matter Phys. \textbf{6}, 15 (2015).}
  \bibitem{SierantRPP2025}{P. Sierant, M. Lewenstein, A. Scardicchio, L. Vidmar, and J. Zakrzewski, Many-body localization in the age of classical computing*, Rep. Prog. Phys. \textbf{88}, 026502 (2025).}  
  \bibitem{BardarsonPRL2012}{J. H. Bardarson, F. Pollmann, and J. E. Moore, Unbounded growth of entanglement in models of many-body localization, Phys. Rev. Lett. \textbf{109}, 017202 (2012).}
  \bibitem{NandkishorePRB2014}{R. Nandkishore, S. Gopalakrishnan, and D. A. Huse, Spectral features of a many-body-localized system weakly coupled to a bath, Phys. Rev. B \textbf{90}, 064203 (2014).}
  \bibitem{GopalakrishnanPRB2016}{S. Gopalakrishnan, K. Agarwal, E. A. Demler, D. A. Huse, and M. Knap, Griffiths effects and slow dynamics in nearly many-body localized systems, Phys. Rev. B \textbf{93}, 134206 (2016).}
  \bibitem{KjallPRL2014}{J. A. Kj\"all, J. H. Bardarson, and F. Pollmann, Many-body localization in a disordered quantum Ising chain, Phys. Rev. Lett. \textbf{113}, 107204 (2014).}
  \bibitem{PalPRB2010}{A. Pal and D. A. Huse, Many-body localization phase transition, Phys. Rev. B \textbf{82}, 174411 (2010).}
  \bibitem{BauerJSM2013}{B. Bauer and C. Nayak, Area laws in a many-body localized state and its implications for topological order, J. Stat. Mech. \textbf{9}, P09005 (2013).}
  \bibitem{SerbynPapiPRL1102013}{M. Serbyn, Z. Papi\'c, and D. A. Abanin, Universal slow growth of entanglement in interacting strongly disordered systems, Phys. Rev. Lett. \textbf{110}, 260601 (2013).}
  \bibitem{SerbynPapiPRL1112013}{M. Serbyn, Z. Papi\'c, and D. A. Abanin, Local conservation laws and the structure of the many-body localized states, Phys. Rev. Lett. \textbf{111}, 127201 (2013).}
  \bibitem{HuseNandkishorePRB2014}{D. A. Huse, R. Nandkishore, and V. Oganesyan, Phenomenology of fully many-body-localized systems, Phys. Rev. B \textbf{90}, 174202 (2014).}
  \bibitem{HamanakaPRB2025}{S. Hamanaka and K. Kawabata, Multifractality of the many-body non-Hermitian skin effect, Phys. Rev. B \textbf{111}, 035144 (2025).}
  \bibitem{RichterPRE2025}{J. Richter, L. S\'a, and M. Haque, Integrability versus chaos in the steady state of many-body open quantum systems, Phys. Rev. E \textbf{111}, 064103 (2025).}
  \bibitem{DalibardPRL1992}{J. Dalibard, Y. Castin, and K. M\o{}lmer, Wave-function approach to dissipative processes in quantum optics, Phys. Rev. Lett. \textbf{68}, 580 (1992).}
  \bibitem{CarmichaelPRL1993}{H. J. Carmichael, Quantum trajectory theory for cascaded open systems, Phys. Rev. Lett. \textbf{70}, 2273 (1993).}
  \bibitem{PlenioRMP1998}{M. B. Plenio and P. L. Knight, The quantum-jump approach to dissipative dynamics in quantum optics, Rev. Mod. Phys. \textbf{70}, 101 (1998).}
  \bibitem{DaleyAP2014}{A. J. Daley, Quantum trajectories and open many-body quantum systems, Advances in Physics \textbf{63}, 77 (2014).}
  \bibitem{GongKawabataPRX2018}{Z. Gong, Y. Ashida, K. Kawabata, K. Takasan, S. Higashikawa, and M. Ueda, Topological phases of non-Hermitian systems, Phys. Rev. X \textbf{8}, 031079 (2018).}
  \bibitem{KawabataPRX2019}{K. Kawabata, K. Shiozaki, M. Ueda, and M. Sato, Symmetry and topology in non-Hermitian physics, Phys. Rev. X \textbf{9}, 041015 (2019).}
  \bibitem{BenderPRL1998}{C. M. Bender and S. Boettcher, Real spectra in non-Hermitian Hamiltonians having PT symmetry, Phys. Rev. Lett. \textbf{80}, 5243 (1998).}
  \bibitem{HatanoNelsonPRL1996}{N. Hatano and D. R. Nelson, Localization transitions in non-Hermitian quantum mechanics, Phys. Rev. Lett. \textbf{77}, 570 (1996).}
  \bibitem{RotterJPAMT2009}{I. Rotter, A non-Hermitian Hamilton operator and the physics of open quantum systems, J. Phys. A: Math. Theor. \textbf{42}, 153001 (2009).}
  \bibitem{AshidaAP2020}{Y. Ashida, Z. Gong, and M. Ueda, Non-Hermitian physics, Advances in Physics \textbf{69}, 249 (2020).}
  \bibitem{WangSutharPRB2023}{Y.-C. Wang, K. Suthar, H. H. Jen, Y.-T. Hsu, and J.-S. You, Non-Hermitian skin effects on thermal and many-body localized phases, Phys. Rev. B \textbf{107}, L220205 (2023).}
  \bibitem{SutharPRB2025}{K. Suthar, Boundary-driven many-body phase transitions in a non-Hermitian disordered fermionic chain, Phys. Rev. B \textbf{111}, 064202 (2025).}
  \bibitem{NandyPathakXianPRB2025}{P. Nandy, T. Pathak, Z.-Y. Xian, and J. Erdmenger, Krylov space approach to singular value decomposition in non-Hermitian systems, Phys. Rev. B \textbf{111}, 064203 (2025).}
  \bibitem{TekurPRB2024}{S. H. Tekur, M. S. Santhanam, B. K. Agarwalla, and M. Kulkarni, Higher-order gap ratios of singular values in open quantum systems, Phys. Rev. B \textbf{110}, L241410 (2024).}
  \bibitem{KawabataPRXQ2023}{K. Kawabata, Z. Xiao, T. Ohtsuki, and R. Shindou, Singular-value statistics of non-Hermitian random matrices and open quantum systems, PRX Quantum \textbf{4}, 040312 (2023).}
  \bibitem{RoccatiPRB2024}{F. Roccati, F. Balducci, R. Shir, and A. Chenu, Diagnosing non-Hermitian many-body localization and quantum chaos via singular value decomposition, Phys. Rev. B \textbf{109}, L140201 (2024).}
  \bibitem{DysonJMP1401962}{F. J. Dyson, Statistical theory of the energy levels of complex systems. I, Journal of Mathematical Physics \textbf{3}, 140 (1962).}
  \bibitem{DysonJMP1571962}{F. J. Dyson, Statistical theory of the energy levels of complex systems. II, Journal of Mathematical Physics \textbf{3}, 157 (1962).}
  \bibitem{GuhrPR1998}{T. Guhr, A. M\"{u}ller--Groeling, and H. A. Weidenm\"{u}ller, Random-matrix theories in quantum physics: common concepts, Physics Reports \textbf{299}, 189 (1998).}
  \bibitem{BeenakkerRMP1997}{C. W. J. Beenakker, Random-matrix theory of quantum transport, Rev. Mod. Phys. \textbf{69}, 731 (1997).}
  \bibitem{PrasadPRB2025}{M. Prasad, S. H. Tekur, B. K. Agarwalla, and M. Kulkarni, Assessment of spectral phases of non-Hermitian quantum systems through complex and singular values, Phys. Rev. B \textbf{111}, L161408 (2025).}
  \bibitem{BaggioliPRD2025}{M. Baggioli, K.-B. Huh, H.-S. Jeong, X. Jiang, K.-Y. Kim, and J. F. Pedraza, Singular value decomposition and its blind spot for quantum chaos in non-Hermitian Sachdev-Ye-Kitaev models, Phys. Rev. D \textbf{111}, L101904 (2025).}
  \bibitem{NandyPathakTezukaPRB2025}{P. Nandy, T. Pathak, and M. Tezuka, Probing quantum chaos through singular-value correlations in the sparse non-Hermitian Sachdev-Ye-Kitaev model, Phys. Rev. B \textbf{111}, L060201 (2025).}
  \bibitem{HamazakiPRL2019}{R. Hamazaki, K. Kawabata, and M. Ueda, Non-Hermitian many-body localization, Phys. Rev. Lett. \textbf{123}, 090603 (2019).}
  \bibitem{ZhaiPRB2020}{L.-J. Zhai, S. Yin, and G.-Y. Huang, Many-body localization in a non-Hermitian quasiperiodic system, Phys. Rev. B \textbf{102}, 064206 (2020).}
  \bibitem{LiuXuPRB2023}{J. Liu and Z. Xu, From ergodicity to many-body localization in a one-dimensional interacting non-Hermitian Stark system, Phys. Rev. B \textbf{108}, 184205 (2023).}
  \bibitem{LiYuPRA2023}{H.-Z. Li, X.-J. Yu, and J.-X. Zhong, Non-Hermitian Stark many-body localization, Phys. Rev. A \textbf{108}, 043301 (2023).}
  \bibitem{SakhrPRE2006}{J. Sakhr and J. M. Nieminen, Spacing distributions for point processes on a regular fractal, Phys. Rev. E \textbf{73}, 036201 (2006).}
  \bibitem{AtasJPAMT2013}{Y. Y. Atas, E. Bogomolny, O. Giraud, P. Vivo, and E. Vivo, Joint probability densities of level spacing ratios in random matrices, J. Phys. A: Math. Theor. \textbf{46}, 355204 (2013).}
  \bibitem{CorpsPRE2020}{\'A. L. Corps and A. Rela\~no, Distribution of the ratio of consecutive level spacings for different symmetries and degrees of chaos, Phys. Rev. E \textbf{101}, 022222 (2020).}
  \bibitem{TekurPRB2018}{S. H. Tekur, U. T. Bhosale, and M. S. Santhanam, Higher-order spacing ratios in random matrix theory and complex quantum systems, Phys. Rev. B \textbf{98}, 104305 (2018).}
  \bibitem{KumarPRB2023}{P. Kumar and R. N. Bhatt, Scaling of entanglement entropy at quantum critical points in random spin chains, Phys. Rev. B \textbf{108}, L241113 (2023).}
  \bibitem{TianPRB2025}{H. Tian, T. He, and X. Wu, Finite-size scaling of half-chain entanglement entropy in the one-dimensional transverse field Ising model and the XX model, Phys. Rev. B \textbf{111}, 104437 (2025).}
  \bibitem{EversPRL2000}{F. Evers and A. D. Mirlin, Fluctuations of the inverse participation ratio at the Anderson transition, Phys. Rev. Lett. \textbf{84}, 3690 (2000).}
  \bibitem{SutharWangPRB2022}{K. Suthar, Y.-C. Wang, Y.-P. Huang, H. H. Jen, and J.-S. You, Non-Hermitian many-body localization with open boundaries, Phys. Rev. B \textbf{106}, 064208 (2022).}
  \bibitem{BertiniPRL2018}{B. Bertini, P. Kos, and T. Prosen, Exact spectral form factor in a minimal model of many-body quantum chaos, Phys. Rev. Lett. \textbf{121}, 264101 (2018).}
  \bibitem{WeiPRE2024}{Z. Wei, C. Tan, and R. Zhang, Generalized spectral form factor in random matrix theory, Phys. Rev. E \textbf{109}, 064208 (2024).}
  \bibitem{BrezinPRE1997}{E. Br\'ezin and S. Hikami, Spectral form factor in a random matrix theory, Phys. Rev. E \textbf{55}, 4067 (1997).}
  \bibitem{SaPRX2020}{L. S\'a, P. Ribeiro, and T. Prosen, Complex spacing ratios: A signature of dissipative quantum chaos, Phys. Rev. X \textbf{10}, 021019 (2020).}
  \bibitem{YusipovAIJNS2022}{I. I. Yusipov and M. V. Ivanchenko, Quantum Lyapunov exponents and complex spacing ratios: Two measures of dissipative quantum chaos, Chaos: An Interdisciplinary Journal of Nonlinear Science \textbf{32}, 043106 (2022).}
  \bibitem{PeronPRE2020}{T. Peron, B. M. F. De Resende, F. A. Rodrigues, L. D. F. Costa, and J. A. M\'endez-Berm\'udez, Spacing ratio characterization of the spectra of directed random networks, Phys. Rev. E \textbf{102}, 062305 (2020).}
  \bibitem{WangSunPRB2021}{Y.-Y. Wang, Z.-H. Sun, and H. Fan, Stark many-body localization transitions in superconducting circuits, Phys. Rev. B \textbf{104}, 205122 (2021).}
  \bibitem{AtasPRL2013}{Y. Y. Atas, E. Bogomolny, O. Giraud, and G. Roux, Distribution of the Ratio of Consecutive Level Spacings in Random Matrix Ensembles, Phys. Rev. Lett. \textbf{110}, 084101 (2013).}
  \bibitem{ZhangZhangPRB2020}{G.-Q. Zhang, D.-W. Zhang, Z. Li, Z. D. Wang, and S.-L. Zhu, Statistically related many-body localization in the one-dimensional anyon Hubbard model, Phys. Rev. B \textbf{102}, 054204 (2020).}
  \bibitem{SuntajsPRB2020}{J. \v{S}untajs, J. Bon\v{c}a, T. Prosen, and L. Vidmar, Ergodicity breaking transition in finite disordered spin chains, Phys. Rev. B \textbf{102}, 064207 (2020).}
  \bibitem{ModakPRL2015}{R. Modak and S. Mukerjee, Many-body localization in the presence of a single-particle mobility edge, Phys. Rev. Lett. \textbf{115}, 230401 (2015).}
  \bibitem{LiProsenPRL2021}{J. Li, T. Prosen, and A. Chan, Spectral statistics of non-Hermitian matrices and dissipative quantum chaos, Phys. Rev. Lett. \textbf{127}, 170602 (2021).}
  \bibitem{GhoshPRB2022}{S. Ghosh, S. Gupta, and M. Kulkarni, Spectral properties of disordered interacting non-Hermitian systems, Phys. Rev. B \textbf{106}, 134202 (2022).}
  \bibitem{HaakeS2010}{F. Haake, Quantum Signatures of Chaos (Springer, Berlin, 2010).}
  \bibitem{CotlerJHEP2017}{J. S. Cotler, G. Gur-Ari, M. Hanada, J. Polchinski, P. Saad, S. H. Shenker, D. Stanford, A. Streicher, and M. Tezuka, Black holes and random matrices, J. High Energy Phys. \textbf{5}, 118 (2017).}
  \bibitem{LiuPRD2018}{J. Liu, Spectral form factors and late time quantum chaos, Phys. Rev. D \textbf{98}, 086026 (2018).}
  \bibitem{LozejE2023}{\v{C}. Lozej, Spectral form factor and dynamical localization, Entropy \textbf{25}, 451 (2023).}
  \bibitem{BianchiJHEP2024}{M. Bianchi, M. Firrotta, J. Sonnenschein, and D. Weissman, From spectral to scattering form factor, J. High Energy Phys. \textbf{6}, 189 (2024).}
  \bibitem{SuntajsPRE2020}{J. \v{S}untajs, J. Bon\v{c}a, T. Prosen, and L. Vidmar, Quantum chaos challenges many-body localization, Phys. Rev. E \textbf{102}, 062144 (2020).}
  \bibitem{PrakashPRR2021}{A. Prakash, J. H. Pixley, and M. Kulkarni, Universal spectral form factor for many-body localization, Phys. Rev. Research \textbf{3}, L012019 (2021).}
  \bibitem{SchulzHooleyPRL2019}{M. Schulz, C. A. Hooley, R. Moessner, and F. Pollmann, Stark many-body localization, Phys. Rev. Lett. \textbf{122}, 040606 (2019).}
  \bibitem{WannierRMP1962}{G. H. Wannier, Dynamics of band electrons in electric and magnetic fields, Rev. Mod. Phys. \textbf{34}, 645 (1962).}
  \bibitem{WannierPR1960}{G. H. Wannier, Wave functions and effective Hamiltonian for Bloch electrons in an electric field, Phys. Rev. \textbf{117}, 432 (1960).}
  \bibitem{PrasadPRB2024}{Y. Prasad and A. Garg, Single-particle excitations across the localization and many-body localization transition in quasiperiodic systems, Phys. Rev. B \textbf{109}, 094204 (2024).}
  \bibitem{JanaPRB2024}{A. Jana, V. R. Chandra, and A. Garg, Universal properties of single-particle excitations across the many-body localization transition, Phys. Rev. B \textbf{109}, 214209 (2024).}

\end{thebibliography}
\end{document}